\documentclass[%
    reprint,
    amsmath,
    amssymb,
    aps,
    pra,
    floatfix,
    superscriptaddress
]{revtex4-2}

\usepackage{graphicx} 
\usepackage{bm} 
\usepackage[colorlinks, linkcolor = black, citecolor = black, filecolor = black, urlcolor = blue]{hyperref}
\usepackage{braket}

\begin{document}

\title{Addressing requirements for crosstalk-free quantum-gate operation in many-body nanofiber cavity QED systems}

\author{Tim Keller}
\email{tim.keller@aoni.waseda.jp}
\author{Seigo Kikura}
\affiliation{Department of Applied Physics, Waseda University, 3-4-1 Okubo, Shinjuku-ku, Tokyo 169-8555, Japan}
\author{Rui Asaoka}
\author{Yasunari Suzuki}
\author{Yuuki Tokunaga}
\affiliation{Computer \& Data Science Laboratories, NTT Corporation, Musashino, Tokyo 180-8585, Japan}
\author{Takao Aoki}
\email{takao@waseda.jp}
\affiliation{Department of Applied Physics, Waseda University, 3-4-1 Okubo, Shinjuku-ku, Tokyo 169-8555, Japan}
\affiliation{RIKEN Center for Quantum Computing (RQC), Wako, Saitama 351-0198, Japan}  
\date{\today}

\begin{abstract}
A distributed network architecture in which flying photons connect individual modules containing stationary atomic qubits is a promising approach for scaling up neutral-atom based quantum-computing platforms.
We consider an all-fiber based platform consisting of nanofiber cavity QED systems interconnected via conventional optical fibers.
Each nanofiber cavity is strongly coupled to multiple atoms through its evanescent field, and atom pairs within one cavity (local) or two distant cavities (remote) are addressed for performing photon-mediated quantum logic gates on them by controlling the effective light-matter coupling via local AC Stark shifts and atom-fiber distance. 
We numerically evaluate the required parameters for achieving nearly crosstalk-free gate operation using these targeting methods by calculating average gate fidelities, success probabilities, and Pauli error rates for both local and remote controlled-$Z$ gates. 
For the case of perfect addressing, we also analytically determine the theoretical optimum gate performance as limited by cavity reflectivity, cooperativity, and qubit level-splitting.  
\end{abstract}

\maketitle

\section{Introduction}
\label{sec:intro}
Neutral atoms are a promising platform for realizing fault-tolerant quantum computation and have seen much progress recently~\cite{graham2022multi,evered2023high,bluvstein2024logical}. 
One approach for scaling up such systems towards useful quantum information processing is a distributed architecture. Multiple smaller computing nodes, in which it is easier to achieve the high levels of control and fidelity necessary for quantum operations, are interlinked to form a large-scale quantum computer \cite{reiserer2015cavity,covey2023quantum,menon2020nanophotonic,main2025distributed}.
\begin{figure}
    \includegraphics[width=\columnwidth]{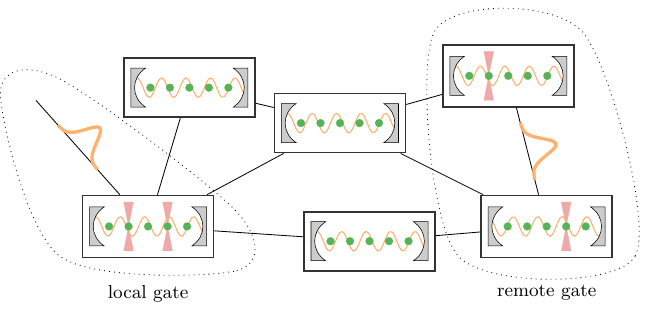}
    \caption{Distributed quantum computing architecture. Individual computing nodes consisting of atomic qubits coupled to cavities are interlinked via conventional optical fibers to form a network in which photon-mediated controlled phase-flip gates can be performed either locally on two atomic qubits targeted within the same cavity or remotely on one atom targeted from two distinct cavities each.} 
    \label{fig:distributed_computing}
\end{figure}
A natural choice for forming such connections in a network is the usage of photons and optical fibers. Placing the atoms at each of the computing nodes in an optical cavity enhances the light-matter interaction and provides the strong coupling required for an efficient connection, similar to the schematic shown in Fig.~\ref{fig:distributed_computing}.
Some of the building blocks for intra- and inter-cavity gate operations have already been demonstrated experimentally, e.g.~local atom-atom controlled-$Z$ (CZ) gates between two atoms inside a single cavity \cite{welte2018photon} or remote CZ gates between two cavities with one atom each \cite{daiss2021quantumlogic}. 

In order to scale up such systems to a larger number of qubits, it is necessary to perform both local and non-local two-qubit gates between arbitrary pairs with high fidelity \cite{xia2015randomized}. In the case of cavity QED systems, it is therefore essential to reliably switch on or off the coupling of individual atoms to the cavity, since all atoms couple to the cavity collectively unless further measures are taken.
Successful demonstrations of targeting and addressing single atoms in setups with multiple atomic qubits per cavity have recently been reported in Refs.~\cite{langenfeld2021quantum,thomas2024fusion,hartung2024quantumnetwork} with two individually addressable atomic qubits in a single cavity for quantum key distribution and photonic graph state fusion.
Similarly, in Ref.~\cite{hu2025site} individual atoms trapped in an optical tweezer array were selectively coupled to a cavity for reading out their quantum state. There, the addressing mechanism is mainly realized through laser beams that locally shift the transition frequency of individual atoms on or off the cavity resonance via the AC Stark effect. In this way, all but the targeted atom are `hidden' from the cavity.  

In this work, we focus on a novel all-fiber based approach, in which the atoms are trapped in optical tweezer arrays next to nanofiber cavities. They interact with the evanescent light surrounding the nanofiber and communicate with distant nanofiber-cavity computing nodes via regular optical fibers \cite{kato2019observation,white2019cavity,hirayama2022hybrid,sunami2025scalable}. This architecture combines strong light-matter coupling with the accessibility of conventional optical fiber networks and is therefore ideally suited for the creation of large-scale distributed quantum computing networks. 
Because the cavity coupling strength depends on the atom-fiber distance, this approach further offers an intrinsic addressing mechanism for gradually controlling the light-matter coupling of individual atoms by moving them to or from the fiber.  
Using both local light shifts and atom-fiber distance as targeting mechanisms, we numerically model a nanofiber cavity network with several atoms per cavity and determine parameter regimes for achieving nearly crosstalk-free gate operation at the example of photon-mediated local and remote CZ gates \cite{Goto2010condition,duan2005robust}.
The gate performance mainly depends on cavity reflectivity and cooperativity, and we provide analytic approximations for both the average and the maximum gate fidelity. In particular, we show that the latter is also limited by the finite qubit level-splitting. 
Our results focus on the limit of long photon pulses whose amplitude varies slowly compared to the cavity time-scales. We briefly discuss the implications of shorter pulses especially with regard to the remote gate performance and how key quantities in our model need to be adjusted in this case.  

The paper is structured as follows. In Sec.~\ref{sec:quantum_gates} we first introduce the general notation and the atomic-state-dependent phase-flip mechanism for a photon reflected from an optical cavity, which is the core building block for cavity QED-based quantum computing. We then introduce the local and remote atom-atom CZ gates based on this principle, as well as a method for calculating their average gate fidelities and success probabilities, which we use as a benchmark for the addressing quality of individual atoms. 
In Sec.~\ref{sec:model} we first describe the physical setup based on nanofiber cavities with ensembles of $^{133}$Cs atoms trapped next to them in optical tweezer arrays and the different addressing mechanisms for controlling the atom-fiber coupling strength studied in this work. We also calculate the wave function amplitudes or reflection coefficients for the different physical loss channels present in the system before detailing how the average gate fidelities are calculated and presenting analytic results for them. 
In Sec.~\ref{sec:results} we derive analytic approximations for the maximum gate performance and present the numerical results obtained for the fidelity and success probability calculations. First for the baseline case of two atoms or perfect addressing in a many-body system, then for the case of adding atoms to the cavities without any targeting measures before finally determining required addressing strengths for achieving baseline performance even in the many-body case and concluding in Sec.~\ref{sec:conclusion}.
\section{Quantum gates in cavity QED systems}
\label{sec:quantum_gates}
In this section we introduce different photon-mediated controlled phase-flip (CPF) gates, which are at the core of performing local and remote two-qubit gates in a nanofiber-based distributed quantum computing architecture. In particular, we introduce the notation, their physical working mechanism, and how to evaluate their performance by calculating average gate fidelities and success probabilities. 
\subsection{Phase-flip mechanism and notation}
The fundamental working principle behind the quantum gates based on cavity QED systems studied in this paper is the phase-flip, i.e.~a sign-change by a factor of $-1$, in the wave function of a photon reflected from a one-sided optical cavity depending on the detuning between the cavity resonance frequency $\omega_c$ and the photon frequency \cite{duan2004scalable,duan2005robust,kimiaee2020cavity}. 
Placing an atom inside the cavity leads to a change in the resonance frequency of the cavity via Rabi splitting, depending on the coupling strength $g$ between the light field and the atom. By choosing the qubit basis states among the energy level structure in the neutral atom in such a way that only one of the states $\ket{0}$ or $\ket{1}$ couples strongly to the cavity via an excited state at frequency $\omega_a$, the photonic phase-flip mechanism can be used to perform atom-photon as well as atom-atom quantum gates. 

For a photon in the long-pulse limit with an envelope that is varying slowly compared to the dynamics time scales determined by $g$, the cavity decay rate $\kappa_r$ and excited state decay rate $\gamma$, i.e.~a photon exhibiting a narrow frequency band centered around $\omega_p$, the reflection coefficient of the system is given by
\begin{equation}
    r(\omega_p) = 1 + \frac{2\kappa_r}{i\left(\omega_p - \omega_c\right) - \kappa_r + \frac{g^2}{i\left(\omega_p - \omega_a\right) - \gamma}} \; ,
    \label{eq:reflection_coefficient}
\end{equation} 
see Appendix \ref{app:reflection_coefficients} for a detailed derivation. 
Focusing on the case of a photon on resonance with the bare cavity frequency $\omega_p = \omega_c$ impinging on the cavity, the phase-flip mechanism becomes apparent when distinguishing two cases. If there is no atom inside the cavity, i.e.~$g=0$, or in the weak-coupling regime $\left|g^2/[i(\omega_c - \omega_a) - \gamma]\right|\ll\kappa_r$, the photon enters the cavity before being reflected from the perfect end mirror, acquiring a phase shift of $\pi$ or a flip in the sign of its wave function according to
\begin{equation}
    r(\omega_c,g\rightarrow 0) \rightarrow -1 \; .
\end{equation}
If on the other hand $\left|g^2/[i(\omega_c - \omega_a) - \gamma]\right|\gg\kappa_r$, the coupling between atom and cavity leads to a change in the cavity resonance frequency according to the Rabi splitting so that the photon cannot enter the cavity anymore and is already reflected from the partially reflective input mirror without a change in its wave function according to
\begin{equation}
    r(\omega_c,g^2 \gg \kappa_r\gamma) \rightarrow 1 \; .
\end{equation}

For describing quantum gates between different atomic qubits mediated via photons, we depict the photons as polarization qubits with horizontal and vertical polarization basis states denoted as
\begin{equation}
    \ket{H} = \begin{pmatrix}
        1\\ 0
    \end{pmatrix}
    \qquad \text{ and } \qquad
    \ket{V} = \begin{pmatrix}
        0\\ 1
    \end{pmatrix} \; .
\end{equation}
In this notation the Jones matrix for a half-wave plate (HWP) rotated by an angle $\theta = \pi/8$ becomes equivalent to a Hadamard gate for the photon \cite{collett2005field}.

For atomic multi-qubit states we use the default little-endian notation with the first qubit $q_1$ taking up the rightmost position inside the state vector $\ket{a_i} := \ket{q_N{\ldots}q_2q_1}$ for $i=1,\ldots,2^N$ and $q_j\in\lbrace 0,1\rbrace$. Since we only consider individual quantum gates throughout this work, without loss of generality, we can always label the atomic qubits such that the first qubit $q_1$ is the control in the controlled atom-photon gate and that qubits $q_1$ and $q_2$ are control and target in the local and remote atom-atom gates. 
The remaining qubits are arbitrarily labeled across the individual computing modules while keeping track of the cavity in which each atomic qubit is placed. 

\subsection{Local atom-atom controlled phase-flip gates}
\begin{figure}
    \centering
    \includegraphics{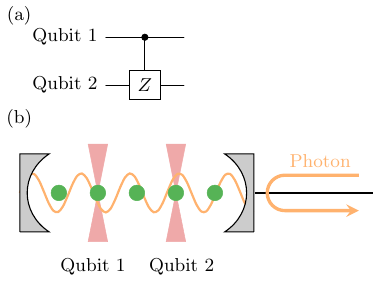}
    \caption{Circuit diagram (a) and sketch (b) of the physical setup for performing a photon-mediated controlled-$Z$ gate between two atomic qubits located in the same cavity and targeted for example via local light shifts from focused laser beams.}
    \label{fig:local_atom_atom_gate}
\end{figure}
For performing an atom-atom CPF or equivalently CZ gate between two atomic qubits inside the same cavity, only a single photon with specific frequency $\omega_p$ is necessary, as shown in Fig.~\ref{fig:local_atom_atom_gate}. Assuming for simplicity that only the control and target atom are placed in the cavity for a total of $N=2$, coupled to it with equal strengths $g_1 = g_2 =:g$, then the reflection coefficient in Eq.~\eqref{eq:reflection_coefficient} is modified according to $g^2\rightarrow Ng^2$.
Denoting the initial general atomic qubit state as $\ket{\psi} = \left(\alpha_2\ket{0} + \beta_2\ket{1}\right)\otimes\left(\alpha_1\ket{0} + \beta_1\ket{1}\right)$ and the photonic state as $\ket{\phi(\omega_p)}_p$, after the reflection the combined atom-photon system state then reads
\begin{equation}
\begin{split}
    &\ket{\psi} = \ket{\phi(\omega_p)}_p\otimes\\[0.5em]
    &(r(N=0)\alpha_2\alpha_1\ket{00} + r(N=1)\alpha_2\beta_1\ket{01} +\\[0.5em]
    &r(N=1)\beta_2\alpha_1\ket{10} + r(N=2)\beta_2\beta_1\ket{11}) \; .
\end{split}
\label{eq:local_gate_final_state}
\end{equation}
By setting the photon frequency to the bare cavity resonance $\omega_p=\omega_c$ and choosing suitable coupling strengths $g$ we have $r(N=0)\rightarrow -1$ and $r(N=1),r(N=2)\rightarrow 1$ so that the desired unitary $\hat{U}=\exp\left(i\pi\ket{00}\bra{00}\right)$ is achieved. 
Similarly, by choosing a photon frequency detuned from the cavity according to the Rabi splitting $\omega_p - \omega_c \sim\sqrt{2}g$ for $N=2$ atoms, we have $r(N=0)$, $r(N=1)\rightarrow 1$, and $r(N=2)\rightarrow -1$, achieving the conventional controlled phase-flip unitary $\hat{U}=\exp\left(i\pi\ket{11}\bra{11}\right)$.
In general, however, we have $|r| < 1$, which indicates photon loss, e.g. through cavity decay or by scattering at the atoms. This results in an unnormalized final state, leading to a reduced gate fidelity and gate success probability as described in Section \ref{subsec:analytics}.
\subsection{Remote atom-atom controlled phase-flip gates}
\begin{figure}
    \centering
    \includegraphics{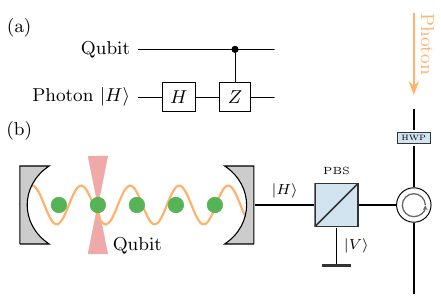}
    \caption{Circuit diagram (a) and sketch (b) of the physical setup for performing an atom-photon controlled-$Z$ gate. A photon initially prepared in $\ket{H}$ is brought into a superposition using a half-wave plate (HWP) acting as a Hadamard gate. A polarizing beam-splitter (PBS) leads to only the $\ket{H}$ component being reflected from the cavity before recombining with the $\ket{V}$ component. Inside the cavity a single atom is targeted, for example via a local AC Stark shift from a focused laser beam. The optical circulator ensures correct order of operation.}
    \label{fig:atom_photon_gate}
\end{figure}
The basic building block for performing photon-mediated two-qubit gates between remote atomic qubits is the atom-photon CPF gate depicted in Fig.~\ref{fig:atom_photon_gate}. 
This sequence, as well as the remote atom-atom variant derived from it in this section, and the local atom-atom gate introduced previously, are also known as Duan-Kimble gate \cite{duan2005robust,raymer2024duankimble}. 
Starting with a photon in $\ket{\phi}_p = \ket{H}$ and a general atomic qubit state $\ket{\psi}_a = \alpha\ket{0} + \beta\ket{1}$, a Hadamard gate on the photon realized via a half-wave plate first leads to $\ket{\phi}_p = (\ket{H} + \ket{V})/\sqrt{2}$. After passing through a polarizing beam splitter (PBS) the $\ket{H}$ component of the photon experiences the atomic-state dependent phase-shift described previously as it is reflected from the cavity before recombining with the $\ket{V}$ component after passing through the PBS again. Note that we are assuming single-photon pulses of sufficient length such that pulse delay and distortion can be neglected \cite{utsugi2022optimal}. The final state of the combined system then reads
\begin{equation}
\begin{split}
    \ket{\psi} = \frac{1}{\sqrt{2}}[r(g=0)&\alpha\ket{H0} + r(g>0)\beta\ket{H1}\\[0.5em] 
     +&\alpha\ket{V0} + \beta\ket{V1}] \; .
\end{split}
\end{equation}
For suitable system parameters we have $r(g=0)\rightarrow -1$ and $r(g>0)\rightarrow 1$ as described earlier and the desired gate unitary $\hat{U}=\exp\left(i\pi\ket{H0}\bra{H0}\right)$ is achieved. Note that, since the phase-flip occurs for atoms in the non-coupling $\ket{0}$ state, there is a difference to conventional CPF gates for which the phase-flip happens in the $\ket{V1}$ state in our notation. However, combining two such gates to a remote atom-atom CZ gate as shown in the following, the final result agrees with the convention again. 
If necessary, a single atom-photon CPF gate can also be made to agree with the conventional notation by switching the PBS output ports and choosing a photon frequency in accordance with the cavity resonance shift due to the Rabi splitting, so that the phase-flip occurs in the $\ket{1}$ state.

\begin{figure}
    \centering
    \includegraphics[width=\columnwidth]{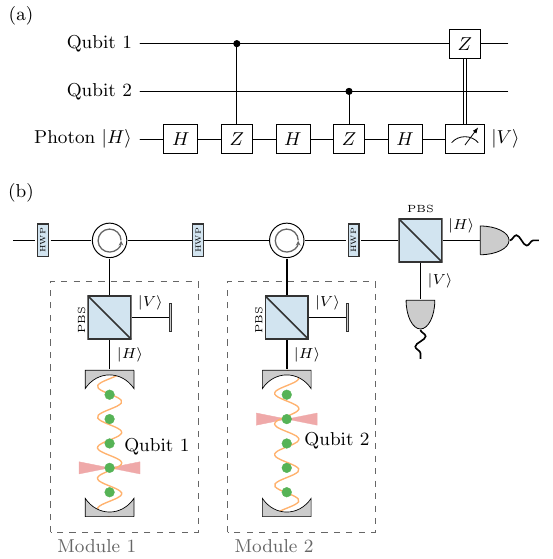}
    \caption{Circuit diagram (a) and sketch (b) of the physical setup for performing a photon-mediated controlled-$Z$ gate between two atomic qubits located in remote computing modules. The photonic qubit is initialized with $\ket{H}$ polarization. Half-wave plates (HWP) act as photonic Hadamard gates and the computing modules consisting of an optical cavity and polarizing beam splitter (PBS) perform atom-photon CZ gates for individually targeted qubits. Optical circulators ensure the correct order of operation. The final $Z$-gate on Qubit 1 is only performed conditional on a $\ket{V}$ polarization measurement outcome.}
    \label{fig:remote_atom_atom_gate}
\end{figure}
Connecting two atom-photon gates as shown in Fig.~\ref{fig:remote_atom_atom_gate} allows performing a controlled phase-flip gate between two remote atomic qubits in different cavities. The initial state of the system is
\begin{equation}
    \ket{\psi} = \ket{H}\otimes\left(\alpha_2\ket{0} + \beta_2\ket{1}\right)\otimes\left(\alpha_1\ket{0} + \beta_1\ket{1}\right)\; .
\end{equation}
After performing the final Hadamard gate on the photon, the state reads
\begin{equation}
\begin{split}
    &\ket{\psi} =\\[0.5em]
    &\frac{\ket{H}}{\sqrt{2}}\left(\alpha_2\alpha_1\ket{00} + \alpha_2\beta_1\ket{01} + \beta_2\alpha_1\ket{10} - \beta_2\beta_1\ket{11}\right)+\\
    &\frac{\ket{V}}{\sqrt{2}}\left(\alpha_2\alpha_1\ket{00} - \alpha_2\beta_1\ket{01} + \beta_2\alpha_1\ket{10} + \beta_2\beta_1\ket{11}\right)
\end{split}
\end{equation}
and the photon polarization is measured. In the case of a $\ket{H}$ measurement outcome, the gate sequence is finished and the desired atom-atom unitary $\hat{U}=\exp\left(i\pi\ket{11}\bra{11}\right)$ is achieved. For a $\ket{V}$ outcome, an additional $Z$-gate must be performed on the first qubit. 
It is important to note that in the physical implementation shown in Fig.~\ref{fig:remote_atom_atom_gate} (b) the final state of the system is modified according to $\ket{H}\rightarrow -\ket{H}$. In other words, in the case of a $\ket{H}$ outcome in the final photon measurement, the targeted atoms acquire an additional phase of $\exp(i\pi)$ since the phase flip in the CZ gates occurs if the respective atom is in the $\ket{0}$ state. This phase needs to be taken into account when performing several gates in a row on different qubits in the distributed architecture.
\subsection{Average gate fidelity, Pauli error rates, and success probability}
A common measure of how well a quantum process approximates a certain desired unitary operation $\hat{U}$ such as a quantum logic gate is the average fidelity \cite{nielsen2010quantum}
\begin{equation}
    F_\mathrm{avg}(\hat{\varepsilon},\hat{U}) = \int d\psi \braket{\psi\vert\hat{U}^\dagger\hat{\varepsilon}\left(\ket{\psi}\bra{\psi}\right)\hat{U}\vert\psi} \; ,
    \label{eq:average_fidelity_int}
\end{equation}
where $\hat{\varepsilon}$ stands for the completely positive and trace preserving (CPTP) map describing the quantum process and the integral over all pure states in the underlying Hilbert space is defined such that $\int d\psi = 1$. Equation \eqref{eq:average_fidelity_int} is generally difficult to evaluate for arbitrary quantum systems. However, the average fidelity can be related to a similar quantity, the entanglement or process fidelity $F_\mathrm{e}(\hat{\varepsilon},\hat{U})$, which is typically easier to compute and describes how well entanglement to a separate system is preserved after applying $\hat{\varepsilon}$ \cite{nielsen2002simple,mayer2018quantum}. 
The entanglement fidelity is defined as
\begin{equation}
\begin{split}
    F_\mathrm{e}(\hat{\varepsilon},\hat{U}) &= \braket{\phi^+\vert\left(\hat{\mathbb{I}}\otimes \hat{U}^\dagger\right)\hat{\chi}(\hat{\epsilon})\left(\hat{\mathbb{I}}\otimes \hat{U}\right)\vert\phi^+}\\[0.5em] 
    &= \operatorname{Tr}\left(\hat{\chi}(\hat{\varepsilon})\hat{\chi}(\hat{U})\right)
\end{split}
\end{equation}
with the identity $\hat{\mathbb{I}}$ and a maximally entangled bipartite state in the originally $d$-dimensional Hilbert space
\begin{equation}
    \ket{\phi^+} = \frac{1}{\sqrt{d}}\sum_{n=1}^d\ket{n}\otimes\ket{n}
\end{equation}
with numbered basis states $\ket{n}$ on which the Choi operator acts according to
\begin{equation}
    \hat{\chi}(\hat{\epsilon}) = \hat{\mathbb{I}}\otimes\hat{\varepsilon}\left(\ket{\phi^+}\bra{\phi^+}\right) \; .
\end{equation}
The connection between average and entanglement fidelity is then given by  
\begin{equation}
    F_\mathrm{avg}(\hat{\varepsilon},\hat{U}) = \frac{dF_\mathrm{e}(\hat{\varepsilon},\hat{U}) + 1}{d + 1} \; .
    \label{eq:average_fidelity}
\end{equation}

Constructing the necessary bipartite states and operators for calculating the entanglement fidelity quickly becomes computationally intense when increasing the number of atomic qubits. We want to note that in our case it is also possible to obtain a good estimate of the average gate fidelity by using the approximation
\begin{equation}
    F_\mathrm{avg}(\hat{\varepsilon},\hat{U}) \approx \bra{a^+}\hat{U}^\dagger\hat{\varepsilon}\left(\ket{a^+}\bra{a^+}\right)\hat{U}\ket{a^+} \; ,
    \label{eq:superposition_input}
\end{equation}
which computes the effect of the quantum process on a trial input state $\ket{a^+} = \sum_{i=1}^{2^N}\ket{a_i}/\sqrt{2^N}$ consisting of an equal superposition of all atomic basis states $\ket{a_i}$.
\begin{figure}
    \centering
    \includegraphics[width=\columnwidth]{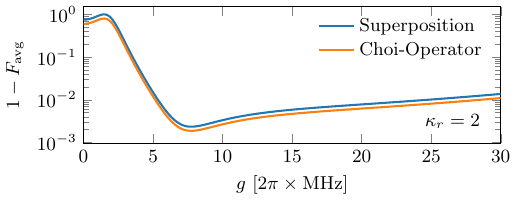}
    \caption{Comparison of average gate fidelity for a post-selected local atom-atom gate with $N=2$ calculated for identical parameters both via the entanglement fidelity in Eq.~\eqref{eq:average_fidelity} and via the superposition input in Eq.~\eqref{eq:superposition_input}. See Sec.~\ref{sec:model} for calculation details.} 
    \label{fig:compare_choi_superpos}
\end{figure}
Figure~\ref{fig:compare_choi_superpos} compares the average fidelities obtained from Eqs.~\eqref{eq:average_fidelity} and \eqref{eq:superposition_input} for a post-selected local atom-atom gate, i.e.~a successful reflection confirmed by photon detection, as a function of the coupling strength $g$ and for a fixed cavity decay rate of $\kappa_r = 2\pi\times 2$ MHz. The calculation details are described in the following Section \ref{sec:model}. 
It shows that both methods yield essentially the same qualitative behavior and that using the approximate method, we generally underestimate the achievable fidelities by some small amount. The approximation therefore provides a way of extending the study to larger systems. Importantly, as shown in Sec.~\ref{subsec:analytics}, we can also use it to obtain analytic expressions in terms of the system parameters for both the local and the remote atom-atom CZ gate.

From a quantum error correction perspective, since the state $\ket{a^+}$ is insensitive to Pauli-$X$ errors, the agreement between both curves indicates that, perhaps unsurprisingly, the photon-mediated CZ gates studied here are biased towards phase-flip errors, similar to other physical quantum computing architectures such as superconducting qubits or trapped ions. This allows for increasing the necessary error threshold in schemes like surface codes for example \cite{aliferis2008faulttolerant,tuckett2019tailoring}.
In a single-qubit quantum channel, this bias $\eta$ is defined as $\eta = p_z/(p_x+p_y) = p_z/(p - p_z)$ for $p = p_x+p_y+p_z$, where $p_i$ is the probability of a Pauli-$i$ error. 
More generally, we can define it as the probability ratio of dephasing errors and non-dephasing errors \cite{puri2020biaspreserving}
\begin{equation}
    \eta = \frac{p_\mathrm{dephasing}}{p_\text{non-dephasing}} \, .
    \label{eq:bias_definition}
\end{equation}
For the two-qubit channels considered in this work we have $p_\mathrm{dephasing} = p_{IZ} + p_{ZI} + p_{ZZ}$ and $p_\text{non-dephasing}$ is the sum of the remaining probabilities except for $p_{II}$, where, for instance, the probability of a $Z$ error occurring in the first qubit and no error (identity) in the second qubit is denoted by $p_{IZ}$. 
We confirm this bias quantitatively by calculating the Pauli error rates for both local and remote gates in Section \ref{sec:results} according to the procedure detailed in Appendix \ref{app:error_rates}.
The Pauli error rates and the entanglement fidelity are further connected via \cite{nielsen2010quantum}
\begin{equation}
    F_\mathrm{e}(\hat{\varepsilon},\hat{U}) = p_{II} = 1 - (p_\mathrm{dephasing} + p_\text{non-dephasing}) \; ,
\end{equation}
where the last equality holds for a trace-preserving quantum channel. In this work we consider a physical model including photon loss, i.e.~a non-trace preserving process, which means that the average fidelity is calculated for successful post-selected events and that the success probability of such events is given by 
\begin{equation}
    p_S = \sum_{ij}p_{ij} < 1 \; ,
    \label{eq:success_probability_error_rates}
\end{equation}
with $i,j\in\lbrace I,X,Y,Z\rbrace$. As shown in the next section, the success probability can also be determined directly from the physical system parameters.
\section{Model for a nanofiber-cavity QED system}
\label{sec:model}
\subsection{Setup and physical addressing mechanisms}
\label{subsec:setup}
\begin{figure}
    \centering
    \includegraphics[width=\columnwidth]{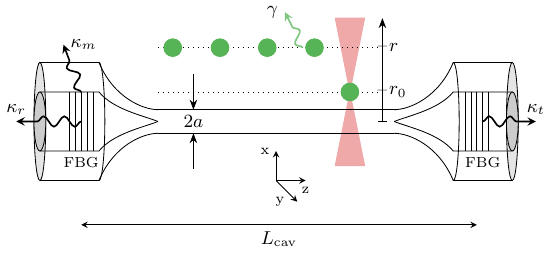}
    \caption{Sketch of a nanofiber cavity that is the basis for the computing modules. The nanofiber portion has a radius of $a$ and at a distance of $L_\mathrm{cav}$ fiber Bragg gratings (FBG) are inscribed into the fiber core, constituting the cavity mirrors with photon reflection, transmission and loss rates labeled $\kappa_r$, $\kappa_t$, and $\kappa_m$, respectively. At a radial distance of $r$ from the fiber center $N$ atoms are placed in a linear configuration via an optical tweezer array. They couple to the guided fiber mode with strengths $g_n$ that may differ for each atom. The targeted atoms are tuned on cavity resonance via a local addressing beam and moved to a fixed distance $r_0$ close to the fiber.}
    \label{fig:setup}
\end{figure}
\begin{figure}
    \centering
    \includegraphics{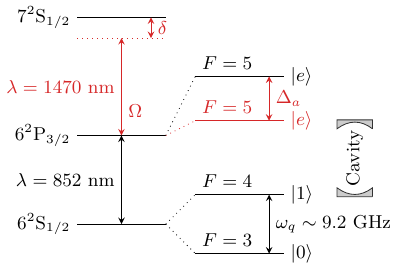}
    \caption{Level structure of $^{133}$Cs. The atomic qubit basis $\lbrace\ket{0},\ket{1}\rbrace$ is encoded in the atomic clock transition with a level splitting of $\omega_q\approx 2\pi\times 9.2$ GHz. The atom is coupled to the cavity via the D2-line transition $\ket{1}\leftrightarrow\ket{e}$ at $\lambda=852.3$ nm and an additional transition at $\lambda = 1470$ nm driven with Rabi frequency $\Omega$ and detuning $\delta$ allows tuning the atomic transition on resonance with the bare cavity frequency via the AC Stark shift for addressing a specific atom.}
    \label{fig:cesium}
\end{figure}

We consider a physical setup of a nanofiber-cavity QED system as shown in Fig.~\ref{fig:setup}. An ensemble of $N$ $^{133}$Cs atoms is trapped close to an optical nanofiber with radius $a$ at a radial distance of $r$ from the fiber center. Fiber Bragg gratings (FBG) inscribed in the fiber core at a distance of $L_\mathrm{cav}$ constitute a one-sided cavity in which the mirror decay rate of the input mirror is much higher than the decay rate of the nearly perfectly reflective end mirror $\kappa_r\gg\kappa_t$. The decay rates can be actively tuned using thermal expansion \cite{kato2015strong}. The atoms undergo spontaneous emission into free space at a rate of $2\gamma$ and they are coupled to the guided fiber mode with strengths $g_n$ that may differ for each atom. Light is assumed to propagate along the $z$-direction and targeted atoms are moved to a fixed position $r_0$ close to the fiber.
In practice, the distance $r_0$ is determined by the minima of the trapping potential generated by the optical tweezer beam interfering with its reflection from the fiber surface \cite{thompson2013coupling}. Similarly, the interference pattern might limit the atom-fiber distance of non-targeted atoms for small values $r\lesssim\lambda_\mathrm{tweezer}$.
The computational basis for the neutral atom qubits is encoded in the states $\ket{0}:=\ket{6 S_{1/2},\; F = 3,\; m_F = 0}$ and $\ket{1}:=\ket{6 S_{1/2},\; F = 4,\; m_F = 0}$ in the level structure of $^{133}$Cs, as shown in Fig.~\ref{fig:cesium}. The transition between these states is commonly used for atomic clocks and has a frequency of $\omega_q\approx 2\pi\times9.1926$ GHz \cite{steck2003cesium,burgardt2023measuring}. The $\ket{1}$ state is coupled to the cavity via the Cesium D2-line transition at $\lambda\approx 852.3$ nm and the excited state $\ket{e}:=\vert 6 P_{3/2},\; F = 5,\; m_F = 0\rangle$. The energy of the $\ket{e}$ state is tunable via an AC Stark shift on an additional transition at $\lambda\approx 1470$ nm. 
The cavity coupling strength as a function of the atomic distance from the fiber is given by \cite{kato2015strong}
\begin{equation}
    g(\mathbf{r}) = \frac{\mathbf{d}\cdot\mathbf{E}(\mathbf{r)}}{\hbar} = \sqrt{\frac{\mu^2\omega}{2\hbar\epsilon_0 V_\mathrm{mode}}}|\phi(\mathbf{r})| \; ,
    \label{eq:cavity_coupling_strength}
\end{equation}
where $\phi(\mathbf{r}) = \vert\mathbf{E}(\mathbf{r})\vert$, the transition frequency is $\omega$ and $\mu = \tilde{\mu}ea_0$ is the transition dipole moment with elementary charge $e$ and Bohr radius $a_0$. 
For the Cesium D2 line the dipole moment is $\tilde{\mu} = 4.4837\times f$ with some factor $f$ depending on the polarization and the Zeeman sublevel $\ket{F;m_F}$. For our choice of $\ket{1}\leftrightarrow\ket{e}$ and $\pi$-polarized light, we have $f = -\sqrt{5/18}$ \cite{steck2003cesium}. 
The mode volume of the dielectric nanofiber cavity is calculated via the cavity length and only for the fundamental $\mathrm{HE}_{11}$ mode according to
\begin{equation}
     V_{\rm mode} = \frac{\int dV \epsilon(\mathbf{r})\,|\mathbf{E}(\mathbf{r})|^2 }{\max [\epsilon(\mathbf{r})\,|\mathbf{E}(\mathbf{r})|^2]} \; ,
\end{equation}
where $\epsilon(\mathbf{r})$ denotes the local relative permittivity. 
Details for calculating the field distribution $\phi_{\mathrm{HE}_{11}}(\mathbf{r})$ of the fundamental fiber mode can be found in Appendix \ref{app:fiber_mode}.

\begin{figure}
    \centering
    \includegraphics{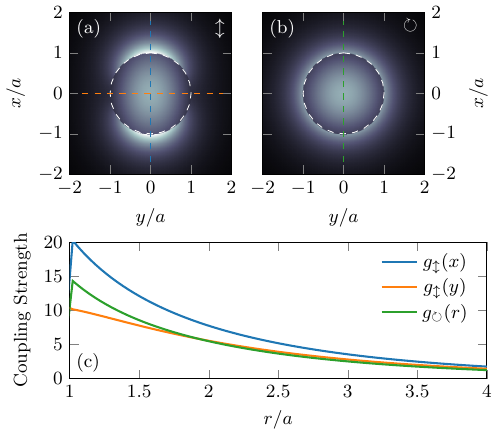}
    \caption{Coupling strength in units of $2\pi\times\mathrm{MHz}$ as a function of atomic distance from fiber, a cavity length of $L_\mathrm{cav}=0.12$ m and radius of $a=200$ nm. For quasi-linearly polarized light along $x$ ($\updownarrow$) the asymmetric intensity distribution outside the fiber (a) leads to different coupling strengths along $x$- and $y$-directions (c). For circularly polarized light ($\circlearrowright$) the radially symmetric intensity distribution (b) results in a coupling strength independent from the azimuthal angle. The color bar in (a) and (b) is normalized to the overall maximum intensity.}
    \label{fig:coupling_strength_pol}
\end{figure}

Figure \ref{fig:coupling_strength_pol} shows the coupling strength for a cavity length of $L_\mathrm{cav}=0.12$ m and a radius of $a=200$ nm as a function of fiber distance for both circularly and quasi-linearly polarized light. Since the axis origin lies in the fiber center, we only show values $r/a>1$ with $r/a=1$ corresponding to an atom placed right on the nanofiber surface. The radially symmetric intensity distribution of circularly polarized light in Fig.~\ref{fig:coupling_strength_pol}(b) results in a coupling strength independent from the azimuthal angle, while the asymmetric intensity distribution for quasi-linearly polarized light in Fig.~\ref{fig:coupling_strength_pol}(a) leads to different coupling strengths along the $x$- and $y$-directions.
Compared to the circularly polarized case, the resulting light-matter coupling is stronger along the polarization axis and weaker along the orthogonal axis. For our purposes it is therefore beneficial to use quasi-linearly polarized light and to place the atoms along the polarization axis. 

In principle, the azimuthal dependence of the coupling strength provides another degree of freedom for controlling the atom targeting quality. But for the fundamental mode HE$_{11}$ and the nanofiber parameters considered here, the asymmetry is not strong enough to provide a viable benefit for moving the atoms by a distance of at least $\sqrt{2}r$ from point $(r,0)$ to point $(0,r)$ compared to simply moving them by the same amount to point $(r+\sqrt{2}r,0)$ along the polarization axis. 
\subsection{Reflection coefficients and loss channels}
The physical setup shown in Fig.~\ref{fig:setup} has several output channels in which an incoming photon can end up. For performing successful quantum gates the photon needs to be reflected from the partially transparent cavity input mirror which has a decay rate of $\kappa_r$. Any other outcome results in photon loss and a failed gate operation. The physically relevant loss channels we are considering here are scattering from the input mirror with rate $\kappa_m$, transmission through the cavity with rate $\kappa_t$ due to imperfections in the fully reflective end mirror and finally scattering at any of the atoms inside the cavity with a rate (see Appendix \ref{app:reflection_coefficients})
\begin{equation}
    \kappa_a = \gamma\sum_{n=1}^N\frac{g_n^2}{\left(\omega_p - \omega_{a,n}\right)^2 + \gamma^2}
\end{equation}
depending on the individual coupling strengths $g_n$, atomic detunings $\Delta_{a,n} = \omega_p - \omega_{a,n}$ and spontaneous emission rate $\gamma$ for each atom. The atomic frequencies $\omega_{a,n}$ can be varied from the unperturbed D2 line at a frequency of $\omega_0$ by locally applying a light shift via the AC Stark effect as in
\begin{equation}
    \omega_{a,n} = \omega_0 - \Delta_\mathrm{AC} \approx \omega_0 - \frac{\Omega^2}{4\delta} \; ,
\end{equation}
where $\Omega$ is the Rabi frequency of the applied laser and $\delta$ is the detuning from this second transition as shown in Fig.~\ref{fig:cesium}.
From a practical point of view, it is also convenient to translate the AC Stark shift into a laser power requirement. Equivalently, for a given laser field $\mathbf{E}$ we have $\Delta_{AC} \approx \alpha(\omega)|\mathbf{E}|^2/4\hbar$ with the dynamic polarizability $\alpha(\omega)$ \cite{grimm2000optical}. 
For a laser with beam waist $w_0$ the average power per unit area and the field strength are related via $P_\mathrm{avg} =  \pi w_0^2 c\epsilon_0|\mathbf{E}|^2/4$, finally resulting in
\begin{equation}
    P_\mathrm{avg} = \pi\hbar c\epsilon_0 w_0^2\frac{\Delta_\mathrm{AC}}{\alpha(\omega)} \; .
    \label{eq:laser_power}
\end{equation}
The dynamic polarizability determining the effective Stark shift of the D2 line is given by the difference in polarizabilities of the involved levels, i.e.~$\alpha(\omega) = \alpha(\omega,6 P_{3/2}) - \alpha(\omega,6 S_{1/2})$, where for linearly polarized light each term can consist of a scalar and a tensor component $\alpha_S$ and $\alpha_T$ respectively. In particular for the excited hyperfine state $\ket{e}=\vert 6 P_{3/2},\; F = 5,\; m_F=0\rangle$ we have 
\begin{equation}
    \alpha(\omega) = \alpha_S(\omega,6 P_{3/2}) - \frac{2}{3}\alpha_T(\omega,6 P_{3/2}) - \alpha_S(\omega,6 S_{1/2}) \; ,
    \label{eq:polarizability}
\end{equation}
where tabulated values of the individual terms as a function of $\omega$ can be found in Ref.~\cite{LeKien2013dynamical}.
We assume that the optical tweezer array trapping the atoms is operated at a magic wavelength of the D2 line at which the differential light shift of the transition vanishes so that the targeting beam alone determines the Stark shift \cite{goban2012demonstration}.
The qubit frequency is not affected within this formalism, since for the $6S_{1/2}$ ground state the tensor polarizability vanishes and all hyperfine states including the computational basis are shifted by an equal amount. However, higher-order effects, such as hyperpolarizability, can also lead to noticeable differential light shifts of the qubit states \cite{carr2016doubly,graham2022multi}.

As shown in Appendix \ref{app:reflection_coefficients}, we can calculate the wave function amplitudes for the remaining loss channels analogously to Eq.~\eqref{eq:reflection_coefficient}. Together with the reflection coefficient they read
\begin{subequations}
    \begin{align}
        r(\omega) &= 1 + \frac{2\kappa_r}{i\left(\omega - \omega_c\right) - \kappa + g(N,\omega)}\; , \\[0.5em]
        t(\omega) &= \frac{2\sqrt{\kappa_r\kappa_t}}{i\left(\omega - \omega_c\right) - \kappa + g(N,\omega)} \; , \\[0.5em]
        m(\omega) &= \frac{2\sqrt{\kappa_r\kappa_m}}{i\left(\omega - \omega_c\right) - \kappa + g(N,\omega)} \; , \\[0.5em]
        a(\omega) &= \frac{2\sqrt{\kappa_r\kappa_a}}{i\left(\omega - \omega_c\right) - \kappa + g(N,\omega)} \; ,
    \end{align}
\end{subequations}
where we have defined the total cavity decay rate $\kappa = \kappa_r + \kappa_m + \kappa_t$ and the overall coupling strength function
\begin{equation}
g(N,\omega) = \sum_{n=1}^N\frac{g_n^2}{i\left(\omega - \omega_{a,n}\right) - \gamma} \; . 
\end{equation}

Following the calculation of Welte et al.~in the appendix of Ref.~\cite{welte2018photon}, we extend their formalism to the case of multiple cavities and several atoms in each cavity. The $2^N$ basis states for a total of $N$ atomic qubits distributed over all cavities are labeled $\ket{a_1} = \ket{00\ldots 0}$, $\ket{a_2} = \ket{00\ldots 1}$,\ldots,$\ket{a_{2^N}} = \ket{11\ldots 1}$. 
For simplicity, we first focus on the case of performing a local atom-atom gate with a single cavity to demonstrate the formalism. The qubits are assumed to be in an arbitrary state with initial density matrix entries $c_{ij}$. After an input photon interacts with the cavity, the density matrix for the final combined state of the atoms and the photon can be written as
\begin{equation}
    \rho_{a,p} = \sum_{i,j=1}^{2^N}c_{ij}\ket{p_i}\bra{p_j}\otimes\ket{a_i}\bra{a_j} \; .
\end{equation}
We consider four distinct outcomes for a photon impinging on the cavity: reflection, transmission, mirror loss, and scattering at the atoms. The photonic states $\ket{p_i}$ are constructed from these four distinct orthogonal modes denoted as $\ket{p_r}$, $\ket{p_t}$, $\ket{p_m}$, and $\ket{p_a}$ respectively, weighed by the previously calculated amplitudes according to
\begin{equation}
    \ket{p_i} = r_i(\omega_p)\ket{p_r} + t_i(\omega_p)\ket{p_t} + m_i(\omega_p)\ket{p_m} + a_i(\omega_p)\ket{p_a} \; ,
\end{equation}
where the individual amplitudes like $r_i$ are evaluated at the coupling strengths $\lbrace g_n\rbrace$ and frequencies $\lbrace \omega_{a,n}\rbrace$ of the respective atomic configuration.
The photonic states are normalized since by construction
\begin{equation}
    \braket{p_i\vert p_i} = \vert r_i(\omega_p)\vert^2 + \vert t_i(\omega_p)\vert^2 + \vert m_i(\omega_p)\vert^2 + \vert a_i(\omega_p)\vert^2 = 1 \; .
\end{equation}

If no post-selection is performed on the system after the local atom-atom gate, we need to trace out the photonic state according to
\begin{equation}
    \rho_{a,\mathrm{out}}^\mathrm{total} = \operatorname{Tr}_p \rho_{a,p} = \sum_{i,j=1}^{2^N}c_{ij}\braket{p_j\vert p_i}\ket{a_i}\bra{a_j} =: \rho_a \circ G_\mathrm{total} \; .
\label{eq:dm_local_total}
\end{equation}
Here $\circ$ denotes the element-wise product and the elements of the matrix $(G_\mathrm{total})_{ij}=\braket{p_j\vert p_i}$ are simply given by the photonic state overlaps. If a reflected photon is detected in $\ket{p_r}$ on the other hand, i.e.~if the operation is post-selected for example by adding an optical circulator and photon detector to the circuit, the final state reads
\begin{equation}
    \rho_{a,\mathrm{out}}^\mathrm{ps} = \frac{\braket{p_r\vert \rho_{a,p}\vert p_r}}{\operatorname{Tr}\left(\braket{p_r\vert \rho_{a,p}\vert p_r}\right)} =: \frac{\rho_{a}\circ G_\mathrm{ps}}{\operatorname{Tr}\left(\rho_a\circ G_\mathrm{ps}\right)}
\label{eq:dm_local_ps}
\end{equation}
instead, where the matrix entries of $(G_\mathrm{ps})_{ij}= r^*_j(\omega)r_i(\omega)$ now only contain reflection coefficients for the different atomic configurations.

For the atom-photon gate shown in Fig.~\ref{fig:atom_photon_gate} the photonic basis states need to be extended to include the vertical polarization component according to
\begin{equation}
    \ket{\tilde{p}_i} = r_i(\omega)\alpha_H\ket{H} + e^{-i\omega\tau_\mathrm{delay}}\alpha_V\ket{V} \; ,
\end{equation}
where $\ket{H}$ corresponds to $\ket{p_r}$ and where $\tau_\mathrm{delay}$ is a complex-valued duration accounting for potential photon loss and pulse delay in the beam splitter arm pertaining to the $\ket{V}$ component. 
The remaining photonic states are left out since only the reflected component is recombined in the beam splitter at the end of the gate. For simplicity, we only consider the case $\tau_\mathrm{delay}\rightarrow 0$ throughout this work, but in particular including a suitable loss can provide another way to optimize gate fidelities at the expense of reduced success probabilities \cite{utsugi2022optimal}. 

For the remote atom-atom gate depicted in Fig.~\ref{fig:remote_atom_atom_gate} only the atoms contained in the respective cavity must be considered when calculating the overall coupling strength $g(N,\omega)$ for each of the two cavity reflections. Just as with the atom-photon gate, the photonic basis states also include both horizontal and vertical components in this case. In order to perform the final conditional $Z$-gate on Qubit 1, we project the density matrix at the end of the gate sequence $\rho_f$ onto both polarization components and only apply an additional $Z$-gate on Qubit 1 in the vertical component before eventually tracing out the photon according to 
\begin{equation}
    \rho_{a,\mathrm{out}}^\mathrm{remote} = \operatorname{Tr}_p\left[\mathcal{P}_H^\dagger\rho_f\mathcal{P}_H + \left(\hat{\sigma}_z^1\right)^\dagger\left(\mathcal{P}_V^\dagger\rho_f\mathcal{P}_V\right)\hat{\sigma}_z^1\right]
    \label{eq:dm_remote}
\end{equation}
with the projection operators $\mathcal{P}_{H/V} = \ket{H/V}\bra{H/V}\otimes\mathbb{I}_{2^N}$. The $Z$-gate on Qubit 1 is denoted as $\hat{\sigma}_z^1=\mathbb{I}_{2^{N-1}}\otimes\hat{\sigma}_z$ and $\mathbb{I}_d$ is the $d$-dimensional identity. Due to the final gate only being applied conditional on the photon measurement outcome, the remote atom-atom gate is automatically post-selected since, ignoring dark counts, a photon detection at the final step signals a successful gate operation.

In general, the quantum gates studied here are successful if the mediating photon is reflected from each of the cavities involved and does not end up in any of the loss channels. The success probability is given by the trace of the system density matrix after applying the gate, so in particular for the local gate we have 
\begin{equation}
    p_S^\mathrm{local} = \operatorname{Tr}\left(\rho_a\circ G_\mathrm{ps}\right) \; ,
\end{equation}
which is also the quantity used to renormalize the density matrix in Eq.~\eqref{eq:dm_local_ps} to account for the post-selection of successful gates.
If the system is in a particular atomic basis state, the success probability of the local gate or photon reflection is simply given by $|r_i(\omega)|^2$. By again using the superposition input $\rho_a = \ket{a^+}\bra{a^+}$ we can then define an average success probability of the quantum channel in terms of system parameters by
\begin{equation}
    p_S^\mathrm{local} = \frac{1}{2^N}\sum_{i=1}^{2^N} |r_i(\omega)|^2 = \frac{\operatorname{Tr}(G_\mathrm{ps})}{2^N}
    \label{eq:ps_local_average}
\end{equation}
for the local CZ gate. For the remote gate we need to consider two reflections with an intermediate photonic Hadamard gate and therefore we have 
\begin{equation}
    p_S^\mathrm{remote} = \frac{\operatorname{Tr}[(\hat{H}_N G_\mathrm{ps}^1 \hat{H}_N^\dagger)\circ G_\mathrm{ps}^2]}{2^{N+1}}
\end{equation}
instead, with the reflection matrices $G_\mathrm{ps}^i$ for both cavities involved and where $\hat{H}_N = \hat{H}\otimes\mathbb{I}_{2^N}$ is the appropriately extended photonic Hadamard gate.
\subsection{Fidelity calculation and analytic results}
\label{subsec:analytics}
\begin{figure}
    \centering
    \includegraphics{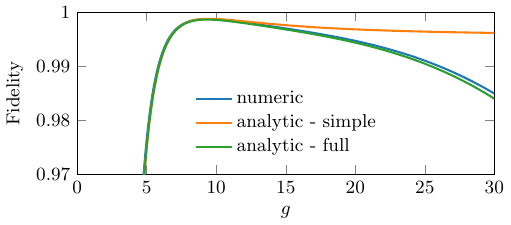}
    \caption{Comparing numerically (Eq.~\eqref{eq:average_fidelity}) and analytically (simplified via Eq.~\eqref{eq:analytic_fidelity_simple} and full via Eq.~\eqref{eq:analytic_fidelity_full}) calculated fidelities of a local atom-atom gate for $N=4$ as a function of coupling strength. The difference between both analytic expressions is due to evaluating the reflection coefficients on atomic resonance and assuming equal coupling strengths. Parameters are $\kappa_r=2.5$, $\kappa_m=\kappa_t=0.1$, $2\gamma = 5.2$, and a detuning of $\Delta_{a}=10^4$ for the non-targeted atoms. All frequencies in units of $2\pi\times\mathrm{MHz}$.}
    \label{fig:analytic_fidelity}
\end{figure}
For both the local and the remote gate, the average gate fidelity is computed numerically by extending the calculations leading to Eqs.~\eqref{eq:dm_local_total}/\eqref{eq:dm_local_ps} and \eqref{eq:dm_remote} to the bipartite system required by the entanglement fidelity according to Eq.~\eqref{eq:average_fidelity}. The maximally entangled state $\ket{\Phi^+}$ is used as the input, and the target unitaries for performing a CZ gate on atoms $n$ and $m$ are $\hat{U}_\mathrm{local} = \exp\left(i\pi\ket{00}_{nm}\bra{00}_{nm}\right)$ and $\hat{U}_\mathrm{remote} = \exp\left(i\pi\ket{11}_{nm}\bra{11}_{nm}\right)$ for the mediating photon on resonance with the bare cavity frequency as explained in Sec.~\ref{sec:quantum_gates}.

In order to also obtain analytic expressions for the fidelity based on system parameters, we can use the previously mentioned superposition input state $\ket{a_+} = \sum_{i=1}^{2^N}\ket{a_i}/\sqrt{2^N}$ as an approximation.
All entries of the initial density matrix are $\left(\rho_{a,\mathrm{in}}\right)_{ij} = 1/2^N$ in this case and for performing a CZ gate on qubits $n$ and $m$ we find
\begin{equation}
\begin{split}
    F &= \braket{a^+\vert\hat{U}^\dagger_\mathrm{local}\frac{G}{\operatorname{Tr}(G)}\hat{U}_\mathrm{local}\vert a^+}\\[0.5em] 
    &= \frac{1}{2^N\operatorname{Tr}(G)}\sum_{i,j=1}^{2^N}G_{ij}\left(1 - 2\delta_{i,A}\right)\left(1 - 2\delta_{j,A}\right) \; ,
    \label{eq:analytic_fidelity_full}
\end{split}
\end{equation}
where $\delta_{i,A}$ is one for any $i$ contained in the set of $\vert A\vert = 2^{N-2}$ indices defined as
\begin{equation}
    A = \left\lbrace 1\leq k \leq 2^N \;\middle\vert\;
    \begin{array}{l}
    \left\lfloor\frac{k-1}{2^{n-1}}\right\rfloor\equiv 0 \pmod{2}\; ,\\[1em] \left\lfloor\frac{k-1}{2^{m-1}}\right\rfloor \equiv 0 \pmod{2}
    \end{array}
    \right\rbrace
\end{equation}
and zero otherwise. $G$ can be either the full matrix $G_\mathrm{total}$ or the post-selected one $(G_\mathrm{ps})_{ij} = r^*_j(\omega)r_i(\omega)$ depending on the physical situation. 

Equation \eqref{eq:analytic_fidelity_full} can be simplified further in the case of a post-selected local CZ gate performed on $N=2$ atoms or equivalently for requiring perfect targeting in the case of $N>2$. On atomic resonance and assuming equal coupling strengths $g_n\equiv g$ the reflection coefficients are simplified to
\begin{equation}
    r(N) = 1 - \frac{2\kappa_r}{\kappa + Ng^2/\gamma}
    \label{eq:reflection_coefficient_simplified}
\end{equation}
and from the full analytic expression in Eq.~\eqref{eq:analytic_fidelity_full} or more simply from Eq.~\eqref{eq:local_gate_final_state} we find
\begin{equation}
    F_\mathrm{local} = \frac{1}{4}\frac{\vert r(0) - 2r(1) - r(2)\vert^2}{\vert r(0)\vert^2 + 2\vert r(1)\vert^2 + \vert r(2)\vert^2} \; .
    \label{eq:analytic_fidelity_simple}
\end{equation}

Figure \ref{fig:analytic_fidelity} compares the analytic fidelities obtained from Eqs.~\eqref{eq:analytic_fidelity_full} and \eqref{eq:analytic_fidelity_simple} for performing a local CZ gate as a function of the coupling strength $g$ with the numeric results. The complete analytic equation \eqref{eq:analytic_fidelity_full} and the numerical results are virtually indistinguishable except for higher values of $g$, where the discrepancy due to the approximation in Eq.~\eqref{eq:superposition_input} gradually grows.
The simplified expression \eqref{eq:analytic_fidelity_simple} leads to slightly larger fidelities for coupling strengths above $g \gtrsim 2\pi\times 10$ MHz. This is due to the stronger addressing requirements that become apparent for larger coupling strengths, as the simplified expression assumes only two atoms or equivalently $N>2$ with perfect targeting.

For calculating the fidelity of the remote gate in a similar fashion, we first determine the state of the photon after the gate sequence in Fig.~\ref{fig:remote_atom_atom_gate}, which reads
\begin{equation}
\ket{\phi}_p = \frac{1}{2\sqrt{2}}
    \begin{pmatrix}
        r_{\mathrm{C2}}(\omega)r_{\mathrm{C1}}(\omega) + r_{\mathrm{C2}}(\omega) + r_{\mathrm{C1}}(\omega) - 1\\
        r_{\mathrm{C2}}(\omega)r_{\mathrm{C1}}(\omega) + r_{\mathrm{C2}}(\omega) - r_{\mathrm{C1}}(\omega) + 1
    \end{pmatrix} \; ,
\end{equation}
where $r_{\mathrm{Ci}}$ denotes the reflection coefficient of cavity $i$. Note that this expression is only valid for sufficiently long photon pulses with negligible pulse delay $\tau_\mathrm{delay}\rightarrow 0$ between the $\ket{H}$ and $\ket{V}$ components as stated earlier. 
Further assuming identical cavities $r_{\mathrm{C1}}(\omega)=r_{\mathrm{C2}}(\omega)$ and using $\ket{a^+}$ as the atomic state as before, we find
\begin{equation}
    F_\mathrm{remote} = \frac{1}{2}\left(F_{\ket{H}} + F_{\ket{V}}\right) 
    \label{eq:remote_gate_analytic_fidelity}
\end{equation}
with the individual terms given by 
\begin{widetext}
\begin{subequations}
    \begin{align}
        F_{\ket{H}} &= \frac{1}{4}\frac{|r(0)^2 + 4r(0) + 2r(1)r(0) - r(1)^2 - 2|^2}{|r(0)^2 + 2r(0) - 1|^2 + 2|r(1)r(0) + r(1) + r(0) - 1|^2 + |r(1)^2 + 2r(1) - 1|^2}\\[0.5em]
        F_{\ket{V}} &= \frac{1}{4}\frac{|r(0)^2 - 2r(0) + r(1)^2 + 2r(1) + 2|^2}{|r(0)^2 + 1|^2 + |r(1)r(0) + r(1) - r(0) + 1|^2 + |r(1)r(0) + r(0) - r(1) + 1|^2 + |r(1)^2 + 1|^2}
    \end{align}
\end{subequations}
\end{widetext}
and using the same function $r(N)$ as in Eq.~\eqref{eq:reflection_coefficient_simplified}. 
\section{Results}
\label{sec:results}
In this section we will first present the system performance for local and remote gates in terms of average gate fidelity, success probability, and Pauli error rates for the baseline case of $N=2$ atoms. We then examine how this performance deteriorates if further atoms are added to the cavity without any addressing or targeting measures before finally determining the strength of the addressing required to achieve the baseline performance even in the many-body case. 

In the following, all rates such as the coupling strength $g$ or the cavity decay rate $\kappa_r$ will be given in units of $2\pi\times\mathrm{MHz}$ unless otherwise indicated. For all results, we always assume a photon on resonance with the cavity frequency and with the atomic transition $\omega_p = \omega_c = \omega_a$, as well as cavity transmission and mirror loss rates of $\kappa_m=\kappa_t=0.1$. The spontaneous emission rate is set to the natural line width of the cesium D2 line $2\gamma = 5.2$ \cite{steck2003cesium}.
\subsection{Baseline case: \texorpdfstring{$N=2$}{N=2} or perfect addressing}
\begin{figure}
    \centering
    \includegraphics[width=\columnwidth]{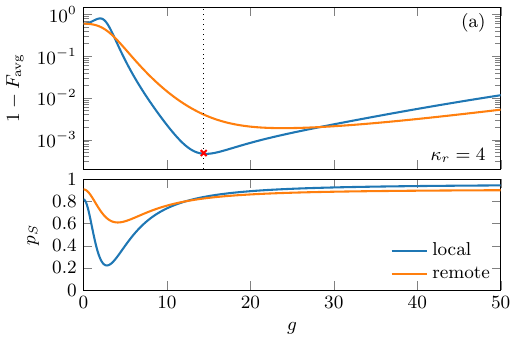}
    \includegraphics[width=\columnwidth]{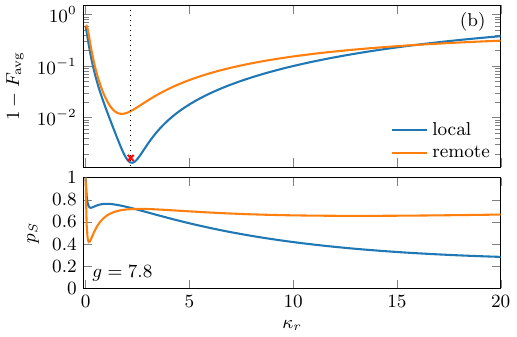}
    \caption{Comparing average infidelity and success probability as a function of coupling strength (a) and cavity decay rate (b) for performing a local or a remote CZ gate for a total of $N=2$ atoms. In all cases we assume equal cavity decay rates, fixed at $\kappa_r = 4$ in panel (a), as well as equal coupling strengths $g_n\equiv g$, fixed at $g=7.8$ in panel (b). The dotted lines and red crosses indicate the approximate positions of maximum fidelity according to Eqs.~\eqref{eq:fidelity_maximum} and \eqref{eq:Fmax_value} respectively.} 
    \label{fig:DK_gate_local_remote_comparison}
\end{figure}
We first examine the case of just two atoms in the system, either both placed in a single cavity for performing a post-selected local CZ gate or control and target qubit placed in distant, but otherwise identical, cavities for the remote protocol. These results serve as the system performance baseline for given parameters and as a benchmark for the necessary addressing strengths in the many-body case studied later on. 

Figure \ref{fig:DK_gate_local_remote_comparison} compares the performance of both versions of the CZ gate in a system of $N=2$ atoms in total or equivalently in a system with perfect addressing in terms of the average gate infidelity $1 - F_\mathrm{avg}$ calculated from Eq.~\eqref{eq:average_fidelity} and the success probability calculated from Eq.~\eqref{eq:success_probability_error_rates}.
In either case, all coupling strengths are assumed to be equal $g_1 = g_2 \equiv g$.
In panel (a) we vary the coupling strength while keeping the cavity decay rate fixed at $\kappa_r = 4$ and in panel (b) the cavity decay rate is varied for fixed coupling strength $g = 7.8$. The chosen values, while somewhat arbitrary, were selected to illustrate the general qualitative behavior of the gate performance in experimentally relevant regimes. They are inspired by previous demonstrations of the strong-coupling regime in similar setups \cite{kato2015strong} and are close to the overall theoretical optimum performance derived in the following. 

\begin{figure}
    \centering
    \includegraphics{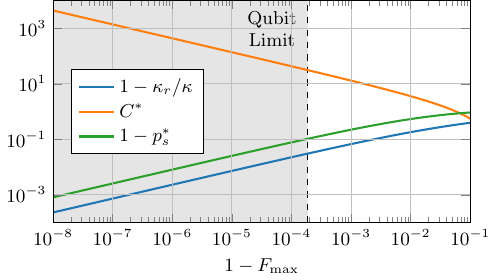}
    \caption{Required loss ratio $\kappa_r/\kappa$ and cooperativity $C^*$ for reaching a certain fidelity $F_\mathrm{max}$ according to Eqs.~\eqref{eq:Fmax_ratio} and \eqref{eq:fidelity_maximum} as well as the gate error probablity $1-p_s^*$ at this point. The shaded area is inaccessible due to the finite qubit level splitting limiting the maximum fidelity according to $1 - F_\mathrm{max}\sim 2\gamma/\omega_q$. Numerically we find $1 - F_\mathrm{max} > 1.88\times10^{-4}$ for the Cesium setup considered here.} 
    \label{fig:local_gate_max_fidelity}
\end{figure}

Since it only involves a single photon reflection, the local gate achieves higher average fidelities over a wide range of experimentally accessible coupling strengths between $g\approx 5 - 30$ with an optimum around $g\approx 15$ for the given cavity parameters. Counterintuitively, the gate performance deteriorates as $g$ increases further and the resulting fidelity maximum is mainly due to two effects.
Firstly, there are other loss channels in the cavity in addition to the semitransparent input mirror, in our case transmission and mirror scattering losses $\kappa_t$ and $\kappa_m$, respectively, which lead to a reduction of the absolute value of the reflection coefficient in the non-coupling state from the ideal value of minus one, $|r(g=0)|<1$.
We can obtain a good estimate for the position of the fidelity maximum from Eq.~\eqref{eq:local_gate_final_state}. The final state of the system will match a renormalized version of the desired target state if $|r(N=0)| = |r(N=1)|$ as long as $|r(N=2)/r(N=1)|\approx 1$. From the first condition, we can derive an approximation for the coupling strength $g_*$ at which the local gate fidelity will become maximum for a photon on resonance according to
\begin{equation}
    C^* := \frac{g^2_*}{\gamma\kappa} = \frac{\kappa_r/\kappa}{1 - \kappa_r/\kappa} - 1 \; ,
    \label{eq:fidelity_maximum}
\end{equation}
as long as $\kappa_r/\kappa>1/2$, which matches closely with the numerically observed value as indicated in Fig.~\ref{fig:DK_gate_local_remote_comparison} and where we have defined the cooperativity $C^*$. Note that $C^* = 2C$ contains an additional factor of two compared to conventional definitions of cooperativity $C$ \cite{reiserer2015cavity}. 
In this way, it is easy to see that the fidelity does not exhibit a local maximum anymore, i.e.~$g_*\rightarrow\infty$, once the influence of the other loss channels vanishes and $\kappa\rightarrow\kappa_r$. At the maximum we further have
\begin{subequations}
    \begin{align}
        F_\mathrm{max} &= 1 - \frac{3\left(1 - \frac{\kappa_r}{\kappa}\right)^2}{4\left[7\left(\frac{\kappa_r}{\kappa}\right)^2 - 4\frac{\kappa_r}{\kappa} + 1\right]} \label{eq:Fmax_value} \; ,\\[0.5em]
        p_s^* &= \frac{7\left(\frac{\kappa_r}{\kappa}\right)^2 - 4\frac{\kappa_r}{\kappa} + 1}{\left(3\frac{\kappa_r}{\kappa} - 1\right)^2}\left(1 - 2\frac{\kappa_r}{\kappa}\right)^2
    \end{align}
\end{subequations}
for the fidelity and success probability. They both depend on the ratio of input-mirror loss to total cavity loss $\kappa_r/\kappa$ but can also be expressed as a function of cooperativity by replacing $\kappa_r/\kappa = (C^*+1)/(C^*+2)$. Equivalently, in order to achieve a certain fidelity $F_\mathrm{max}$, the following loss ratio
\begin{equation}
    \frac{\kappa_r}{\kappa} = 1 - \frac{20\sqrt{F_\mathrm{max}(1-F_\mathrm{max})} + \sqrt{48}(1-F_\mathrm{max})}{\sqrt{75} + 28\sqrt{F_\mathrm{max}(1-F_\mathrm{max})}} 
    \label{eq:Fmax_ratio}
\end{equation}
in combination with the corresponding cooperativity $C^*$ must be reached.
Figure \ref{fig:local_gate_max_fidelity} shows the loss ratio and cooperativity necessary for a certain $F_\mathrm{max}$, as well as the error probability $1 - p_s^*$ at this point. 
If we evaluate the expression for the remote gate fidelity in Eq.~\eqref{eq:remote_gate_analytic_fidelity} at the condition $r(1) = -r(0)$ that maximizes local gate fidelity, it simplifies to
\begin{equation}
    F_\mathrm{remote} \rightarrow \frac{1}{4}\frac{\left(r(0) - 1\right)^4}{\left(r(0)^2 + 1\right)^2} \; .
\end{equation}

The second effect leading to the appearance of a local maximum in the fidelity is the fact that the energy difference $\hbar\omega_q$ between the computational basis states $\ket{0}$ and $\ket{1}$ is large but finite. Very large coupling strengths shift the cavity resonance frequency to such an extent that the otherwise sufficiently detuned and therefore essentially non-coupling $\ket{0}$ state starts interacting with the cavity mode as well, leading to a further reduction in gate fidelity below the maximum value.
Since this effect is stronger for the local gate with two atoms coupled to the cavity, the remote gate eventually obtains a small performance advantage in this ultra-strong coupling regime.
In particular the reflection coefficient $r(0)$ of the non-coupled state is modified according to
\begin{equation}
    r(0) \rightarrow r(0) + 2\frac{\kappa_r}{\kappa}\left[1 - \frac{1}{1+i\frac{2\gamma}{\omega_q}\left(\frac{\kappa_r/\kappa}{1 - \kappa_r/\kappa} - 1\right)}\right] \; ,
\end{equation}
when evaluated at the maximum condition in Eq.~\eqref{eq:fidelity_maximum}. Therefore, we find that, in leading order, the achievable fidelity is in good approximation limited to $1 - F_\mathrm{max}\sim 2\gamma/\omega_q$ by the finite qubit splitting. 
For simplicity, we have assumed so far that both the $\ket{0}$ and the $\ket{1}$ state couple to the cavity via the same excited state $\ket{e}$. In reality, for Cesium qubits the $\ket{0}$ state couples to the cavity through the state $\ket{e'}:=\vert 6 P_{3/2},\; F = 4,\; m_F = 0\rangle$ for $\pi$-polarized light. Therefore, we need to take into account that according to Eq.~\eqref{eq:cavity_coupling_strength} the cavity coupling strength is decreased by a factor of $|\mu_{\ket{0}}/\mu_{\ket{1}}| = \sqrt{3/7}$ and that the effective detuning changes according to $\omega_q' = \omega_q - 2\pi\times 251.1$ MHz \cite{steck2003cesium}. 
Including these effects, we numerically determined the constant of proportionality to be $1 - F_\mathrm{max} \approx 0.7528\times|\mu_{\ket{0}}/\mu_{\ket{1}}|^2\times2\gamma/\omega_q'$ as indicated by the shaded area in Fig.~\ref{fig:local_gate_max_fidelity}. 
This means that for performing a local gate with the Cesium setup considered here, the optimum fidelity of $F_\mathrm{max} = 0.9998$ is reached for a cavity with a loss ratio of $\kappa_r/\kappa = 0.9696$ and a cooperativity of $C^* = 30.86$, close to the parameters used in Fig.~\ref{fig:DK_gate_local_remote_comparison} (a). 

In general, this comparison between local and remote gate performance also touches upon an interesting aspect of a large-scale distributed quantum computing architecture, which is the question of qubit arrangement for a given computational task and system parameters. In other words, if it is beneficial to place two qubits in the same or in different computing modules if, for example, many CZ gates need to be performed on them.
Furthermore, the local gate exhibits a regime around small $g\approx 2.5$ in which the fidelity drops to zero as a result of a zero crossing occurring in the reflection coefficients, which for a single atom on resonance occurs around $\kappa_r = g^2/\gamma$ as can be seen, e.g.,~from Eq.~\eqref{eq:reflection_coefficient}. Due to the `mixing' of the polarization components by the intermediate photonic Hadamard gates, this regime is absent in the remote gate realization. 
Similarly, the local gate generally achieves higher fidelities for most cavity decay rates with another pronounced optimum around $\kappa_r\approx 2.5$ for the given parameters, which agrees well with the prediction from Eq.~\eqref{eq:fidelity_maximum}.

The gate success probabilities mainly depend on the reflection coefficients, and as such they also exhibit a pronounced dip around $g\approx 2.5$ for the given parameters, where the reflection coefficients vanish as mentioned earlier. Except for this minimum, they generally increase as a function of the atomic coupling strengths for fixed cavity decay rate. This is because for the system of $N=2$ atoms the reflection coefficients of three out of four atomic basis states and therefore according to Eq.~\eqref{eq:ps_local_average} also the success probability increase as a function of $g$.
In contrast, for fixed coupling strengths, starting from perfect reflections with $p_S\rightarrow 1$ for $\kappa_r\rightarrow 0$,  the success probability generally decreases with increasing cavity decay rates due to a smaller portion of the incoming photon being reflected at the input mirror. For the remote gate, however, there is almost no effect again because of the intermediate Hadamard gates `mixing' the photon polarization components. 
The dip occurring around small $\kappa_r\approx 0.2$ stems from the reflection coefficient of a non-coupling state vanishing around $\kappa = 2\kappa_r$ and as there are two non-coupling atomic basis states for each cavity in the remote protocol versus a single non-coupling state in the local gate, the dip is more pronounced for the remote gate. 

\begin{figure}
    \centering
    \includegraphics{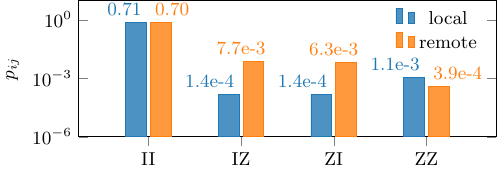}
    \caption{Dephasing Pauli error rates for performing a local or a remote CZ gate on $N=2$ atoms for $g=7.8$ and $\kappa_r=2.5$. See Sec.~\ref{sec:model} and App.~\ref{app:error_rates} for calculation details. All non-dephasing error rates are zero, resulting in an infinitely-biased noise model.}
    \label{fig:pauli_error_rates}
\end{figure}
As mentioned in Sec.~\ref{sec:quantum_gates}, the gate success probability can be decomposed into individual Pauli error rates to further characterize the type of errors. In Fig.~\ref{fig:pauli_error_rates} we perform this decomposition as described in Appendix \ref{app:error_rates} for both the local and the remote gate and exemplary parameters of $g=7.8$ and $\kappa_r=2.5$. 
The largest contributions are from $p_S\sim p_{II}\sim 0.7$, in which case no error occurs. This is followed by dephasing errors $p_{IZ}$, $p_{ZI}$, and $p_{ZZ}$ that are on the order of $10^{-3}$ and $10^{-4}$. 
All other error rates, which make up the non-dephasing errors, are zero within numerical machine precision and not shown in the figure. Therefore, since $p_\text{non-dephasing}\rightarrow 0$ the gates studied in this work are infinitely biased with $\eta\rightarrow\infty$ according to Eq.~\eqref{eq:bias_definition}.
For the local gate, a Z error occurring in both qubits simultaneously, $p_{ZZ} = 1.1\times 10^{-3}$, is about ten times more likely than individual Z errors, which are equally likely for either qubit due to symmetry reasons, $p_{IZ} = p_{ZI} = 1.4\times 10^{-4}$. For the remote gate, individual Z errors $p_{IZ} = 7.7\times 10^{-3} \neq 6.3\times 10^{-3} = p_{ZI}$ are an order of magnitude more likely than simultaneous errors $p_{ZZ} = 3.9\times 10^{-4}$ instead, and they also exhibit an asymmetry since infidelities or errors produced during the first cavity reflection propagate to the second one. 
\subsection{Many-body case without addressing}
In order to demonstrate and quantify the need for qubit targeting, next we perform CZ gates in a many-body system but without any addressing measures applied to the target atoms. For ensuring a fair comparison between local and remote gates and in line with the question of qubit arrangement discussed in the previous section, we focus on systems with a fixed number $N$ of total atoms, which are either all placed in a single cavity when performing a local CZ gate on two of them or which are added evenly to two identical cavities initially only containing the target and control qubit in the remote case. Atoms are added in an alternating fashion starting with the first cavity. For example, for $N=3$ atoms we have two atoms in the first cavity and one atom in the second cavity, and for $N=4$ we have two atoms in each cavity. It is interesting to note that for odd values of $N$ the remote gate performance depends on the qubit distribution and that there is a slight but measurable difference in the resulting fidelities if atoms are added to the second cavity first, i.e.~having one atom in the first and two atoms in the second cavity for $N=3$, similar to the asymmetric Pauli error rates observed in the previous section.
\begin{figure}
    \centering
    \includegraphics{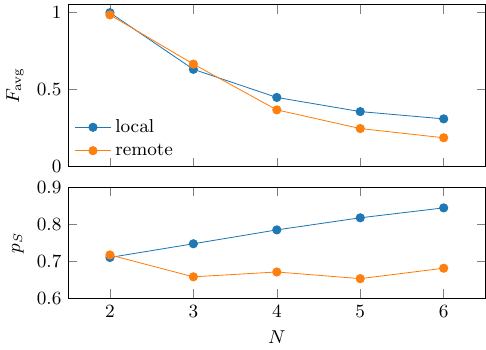}
    \caption{Change in average gate fidelity and success probability for increasing the number of atoms without any addressing mechanisms. Cavity decay rate is $\kappa_r = 2.5$ and all atoms have equal cavity coupling strengths $g_n\equiv 7.8$. Note that despite the increase in success probability for the local gate, the corresponding average gate fidelities are too low to be of practical use, as detailed in the main text.}
    \label{fig:fidelity_ps_local_gate_manybody}
\end{figure}

Figure \ref{fig:fidelity_ps_local_gate_manybody} shows how adding atoms to the system at equal coupling strengths and without performing any addressing at all deteriorates the system performance and leads to a decay in the average gate fidelity. Unsurprisingly, adding even just a single atom for a total of $N=3$ reduces the fidelity to values $F_\mathrm{avg}\sim 0.6$ far below any threshold necessary for practical applications, underscoring the need for qubit targeting mechanisms in our setup. 
Counterintuitively, the success probability of the local gate increases as more atoms are added to the system. A larger number of atoms coupled to the cavity with equal strengths simply means that now on average there are more atomic state configurations for which an enhanced Rabi splitting shifts the cavity resonance frequency further away from its bare value that the incoming photon is tuned to. This results in an on average higher chance for the incoming photon being reflected from the input cavity mirror without any phase shift and therefore the resulting success probability according to our definition Eq.~\eqref{eq:ps_local_average} is increased.
Nevertheless, despite the higher probability of registering a successful gate operation in this case, the resulting fidelities are far too low to be of any use, as mentioned above.
For the remote gate, the success probability largely remains constant as more atoms are added to the system, which is likely again due to the `mixing' effect of the intermediate photonic Hadamard gate.

\begin{figure}
    \centering
    \includegraphics{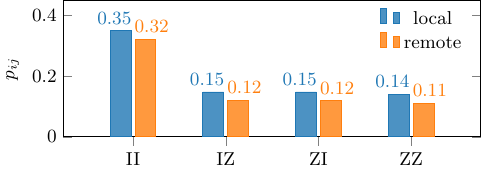}
    \caption{Dephasing Pauli error rates for performing a local or a remote CZ gate on a total of $N=4$ atoms without any addressing for $g=7.8$ and $\kappa_r=2.5$. See Sec.~\ref{sec:model} and App.~\ref{app:error_rates} for calculation details. All non-dephasing error rates are zero, resulting in an infinitely-biased noise model. Note the linear scale compared to Fig.~\ref{fig:pauli_error_rates}.}
    \label{fig:pauli_error_rates_N4_without_addressing}
\end{figure}

The Pauli error rates for the same exemplary case with equal coupling strengths of $g_n\equiv 7.8$, cavity decay rate $\kappa_r=2.5$ and without any addressing are shown in Fig.~\ref{fig:pauli_error_rates_N4_without_addressing} for $N=4$. As before, only the dephasing error rates $p_{IZ}$, $p_{ZI}$ and $p_{ZZ}$ are nonzero and therefore they are still representing an infinitely-biased noise model. However, due to the lack of qubit targeting, they now have the same order of magnitude as $p_{II}$ with comparable error rates for both the local and remote gate.
\subsection{Addressing requirements}
Finally, in this section we want to determine how strongly the effective cavity coupling of the non-targeted atoms has to be suppressed in order to restore the baseline performance and translate this to addressing requirements for the control and target qubit in the physical setup presented in Subsection \ref{subsec:setup}. 

\begin{figure}
    \centering
    \includegraphics{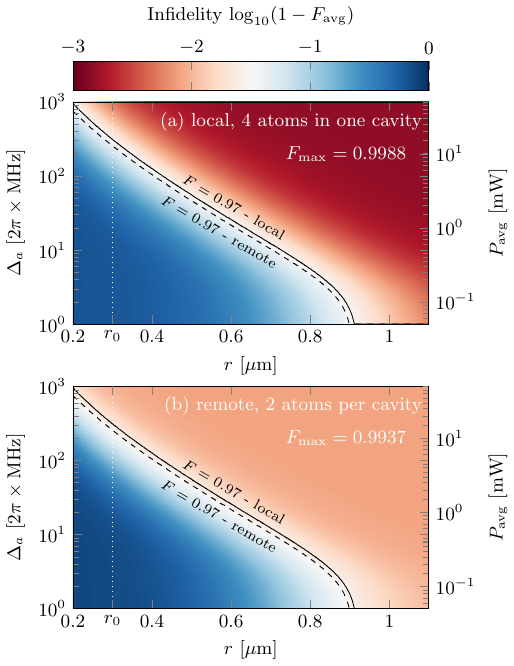}
    \caption{Logarithmic plot of the infidelity $\log_{10}(1-F_\mathrm{avg})$ for performing a local (a) or a remote (b) CZ gate in a system with a total of $N=4$ atoms each (2 per cavity in the remote case) as a function of fiber distance and atomic detuning for the non-targeted atoms. Both target atoms are always kept on resonance $\Delta_{a,1/2}=0$ and at a fixed distance of $r_0=300$ nm where $g_n\approx 10.8$ for quasi-linearly polarized light. The cavity decay rate is $\kappa_r = 2.5$. The laser power $P_\mathrm{avg}$ required for the detuning is calculated for a Gaussian beam with a wavelength of $\lambda=1480$ nm and a beam waist of $w_0=2$ $\mu$m.}
    \label{fig:local_DK_gate_N4_distance_detuning}
\end{figure}
Figure \ref{fig:local_DK_gate_N4_distance_detuning} shows a logarithmic plot of the infidelity after performing (a) local and (b) remote CZ gates on a total of $N=4$ atoms as a function of atomic detuning $\Delta_a$ and fiber distance $r$ of the non-targeted atoms. In the local version, all four atoms are placed in one cavity while they are split equally between two cavities in the remote version. The second y-axis shows the laser power required for achieving the detuning calculated for a Gaussian beam that is red-detuned from the transition shown in Fig.~\ref{fig:cesium} with a wavelength of $\lambda=1480$ nm and a beam waist of $w_0=2$ $\mu$m.
From Ref.~\cite{LeKien2013dynamical} and Eq.~\eqref{eq:polarizability} the polarizability at this wavelength is found to be $\alpha(\lambda = 1480\text{ nm}) \approx 26709.98 \text{ a.u.}$ in atomic units. 
According to Eq.~\eqref{eq:laser_power}, achieving the maximum detuning of $\Delta_a/2\pi = 10^3$ MHz considered here requires a laser power of $50.19$ mW for these parameters, corresponding to a detuning of $19.92$ MHz per mW of laser power. These values are only meant as a rough estimate, in particular since strictly speaking Eq.~\eqref{eq:polarizability} is only valid if the Stark shift is small compared to the hyperfine splitting of the corresponding state. The required laser powers can be reduced further when increasing the polarizability by tuning the addressing beam closer to the atomic transition at $\lambda=1470$ nm. However, as this also increases the scattering rate, leading to heating and decoherence in the atom and potentially reducing the gate fidelity, such a trade-off needs to be considered carefully.

The targeted atoms are always tuned on resonance $\Delta_a=0$ and kept at a fixed distance of $r_0=300$ nm where the cavity coupling strength is $g\approx 10.8$ for quasi-linear polarization, a fiber radius of $a=200$ nm and a cavity length of $L_\mathrm{cav} = 0.12$ m. As before, the cavity decay rate is set to $\kappa_r = 2.5$.
We can see that targeting through detuning alone requires very large AC Stark shifts of $\Delta_a\gtrsim 300$ to achieve fidelities of at least $F>0.97$. On the other hand, placing non-targeted atoms at a distance of $r\gtrsim  1$ $\mu$m eliminates the need for additional addressing via detuning altogether, i.e.~targeted atoms have to be moved by an amount of $r - r_0\gtrsim 0.7$ $\mu$m.
Note that these results are independent of the cavity length. Since we are assuming photons in the long-pulse limit, reflection coefficients, and therefore also quantities like the fidelity calculated from them, depend only on the cavity cooperativity. The cooperativity $C=g^2/2\kappa\gamma$ in turn is invariant under changes in cavity length since its components scale as $g\sim1/\sqrt{L_\mathrm{cav}}$ and $\kappa\sim 1/L_\mathrm{cav}$ \cite{reiserer2015cavity,utsugi2022optimal}.

We further note that our choice of presenting results by combining both targeting methods is mainly intended as a way of gaining an understanding of their relative influence in controlling the atom-cavity coupling strength in the given setup. 
In theory, by combining both methods, the addressing requirements can be reduced significantly, e.g.~the detuning can be decreased by two orders of magnitude to a moderate value of $\Delta_a\gtrsim 10$ when moving the atoms by roughly half the distance of $r - r_0\gtrsim 0.5$ $\mu$m from a position at $r = 0.8$ $\mu$m for the same threshold of $F>0.97$.
From an experimental point of view, however, it is often simpler to adhere to a single targeting mechanism instead. For example, except for duration, there is essentially no practical difference between moving atoms by one or several $\mu$m and therefore one could simply store the atoms far outside the interaction zone and forego the targeting laser entirely. 
Nonetheless, our model can also serve as a general framework for exploring other targeting mechanisms for controlling the collective cavity coupling $g(N,\omega)$. Beyond the natural choices of local AC Stark shifts and atom-fiber distance, it is also conceivable to use techniques such as shelving qubits in separate hyperfine sublevels away from the cavity transition, which have already been successfully demonstrated in neutral-atom computing for mid-circuit readout of individual qubits \cite{graham2023midcircuit,lis2023midcircuit}.

As seen in the baseline results, even with perfect targeting, the remote gate fidelity does not reach the maximum fidelity of the local gate for most parameters, that is $F^\mathrm{remote}\leq F^\mathrm{remote}_\mathrm{max} = 0.9937 < 0.9988 = F^\mathrm{local}_\mathrm{max}$ for the given example. 
However, since for $N=4$ only one atom per cavity has to be decoupled for the remote gate compared to two atoms for the local gate, the remote gate has the advantage that it typically requires less addressing than the local gate for reaching the same fidelity below $F^\mathrm{remote}_\mathrm{max}$ as indicated by the threshold curve of the remote gate lying below the local one for $F=0.97$ as shown in Fig.~\ref{fig:local_DK_gate_N4_distance_detuning}. 
In other words, in the distributed quantum computing architecture studied in this work, loosened addressing requirements can make it more economical to perform a remote CZ gate compared to a local one for applications that do not require achieving $F^\mathrm{local}_\mathrm{max}$.

\begin{figure}
    \centering
    \includegraphics{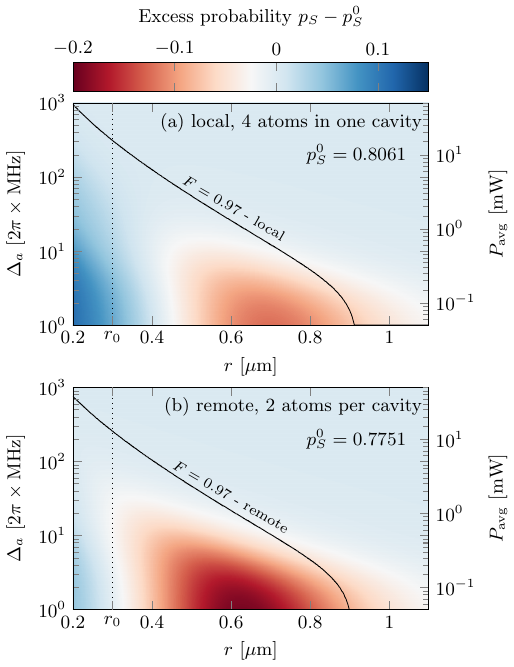}
    \caption{Excess success probability $p_S - p_S^0$ for performing a local (a) or a remote (b) CZ gate in a system with a total of $N=4$ atoms each (2 per cavity in the remote case) as a function of fiber distance and atomic detuning for the non-targeted atoms. Both target atoms are always kept on resonance $\Delta_{a,1/2}=0$ and at a fixed distance of $r_0=300$ nm (dotted line) where $g_n\approx 10.8$ for quasi-linearly polarized light. The excess probability is with regard to the baseline values of $p_S^0=0.8061$ (local) and $p_S^0=0.7751$ (remote) in the case of $N=2$ for identical coupling strengths and cavity decay rates of $\kappa_r=2.5$. The laser power $P_\mathrm{avg}$ required for the detuning is calculated for a Gaussian beam with a wavelength of $\lambda=1480$ nm and a beam waist of $w_0=2$ $\mu$m.}
    \label{fig:ps_local_remote_gate_N4_addressing}
\end{figure}
In Fig.~\ref{fig:ps_local_remote_gate_N4_addressing} we perform the same analysis for the success probability $p_S$ in the case of $N=4$. Since the absolute values of $p_S$ for the local and remote gate are slightly different, we instead plot the excess success probability $p_S - p_S^0$ for easier comparison. Here, $p_S^0$ denotes the success probability in the baseline case of $N=2$ evaluated for the same coupling strengths of $g_n\equiv 10.8$ and cavity decay rates of $\kappa_r=2.5$. We have $p_S^0=0.8061$ and $p_S^0=0.7751$ for the local and the remote gate, respectively.
We find that for both gates the success probability decreases compared to the baseline case mainly in the regime $F<0.97$, where the system does not reach the chosen threshold fidelity. However, especially for the local gate there are also areas of increased success probability at very close fiber distances as a result of the large coupling strengths as explained earlier, as well as a large parameter regime in which the success probability is unchanged despite the reduced fidelity. 
If targeting is controlled mainly via fiber distance and only small detunings of $\Delta_a\approx 1 - 10$, the success probability remains reduced even beyond the chosen fidelity threshold at distances of $r\gtrsim 1$ $\mu$m.  
\section{Conclusion \& Outlook}
\label{sec:conclusion}
We have studied the performance of photon-mediated local and remote CZ quantum gates in a distributed network architecture consisting of several atomic qubits coupled to an optical cavity at each network node. 
We established the system performance in the baseline case of $N=2$ atoms by calculating the average gate fidelity, gate success probability, and Pauli error rates, including several possible loss channels for the mediating photon into the model. We provided approximate analytic expressions for average and maximum fidelities in some cases and showed how these quantities are limited by cavity reflectivity, cooperativity, and in particular qubit level splitting.
We confirmed that these gates exhibit infinitely biased noise models consisting purely of dephasing errors, even in the case where more atoms are added to the cavity without any addressing measures and where the gate fidelity rapidly deteriorates as a result. 

At the example of an all-fiber based setup consisting of $^{133}$Cs qubits trapped close to nanofiber cavities by an optical tweezer array and connected through conventional fibers, we further determined the required addressing strengths for the control and target qubits in order to restore the baseline performance if more than two atoms are coupled to the cavities. 
In particular, we showed that by combining two addressing mechanisms for selectively enhancing the cavity coupling of targeted atoms, namely changing their atom-fiber distance and locally tuning them on cavity resonance via an AC Stark shift, at least in theory the addressing requirements could be reduced significantly compared to just using either of the mechanisms alone.

We further found that the presence of additional atomic qubits can pose interesting questions about qubit arrangement in a network structure. Although generally the local gate can achieve higher maximum fidelities in the baseline case, the addressing requirements for achieving a certain fidelity value below these maximum fidelities were typically lower for the remote gate. Therefore, in certain situations, it can be more desirable or economical to perform a remote gate. 

These considerations are based on the performance calculated purely from the quality of the cavity reflections. For simplicity we have assumed that the additional optical components required for the remote gate, such as half-wave plates, polarizing beam-splitters, optical circulators or photon detectors, as well as the conditional single-qubit Z-gate are working perfectly without incurring any photon loss or infidelities. However, it would be no problem to incorporate imperfections in these components into our model as well for more realistic estimates.
Another source of possible crosstalk that we have not addressed is the finite beam-waist of the lasers creating the local AC Stark shift for bringing targeted atoms close to resonance with the cavity. Typically, they can be focused to radii slightly larger than the wavelength \cite{barredo2018synthetic}, so that even for very large system sizes on the order of $1000$ atoms trapped next to a short cavity of length $L_\mathrm{cav}=3$ mm the effect of unintentionally targeting neighboring atoms is probably negligible for the average atomic distance of $3$ $\mu\mathrm{m}$.

A possible aspect of future work could be the consideration of gate operation times, for example including the effects of photon pulses with finite duration such as pulse delay and distortion, similar to previous studies \cite{asaoka2021requirements,utsugi2022optimal}. In the long-pulse limit used throughout this work, the reflection coefficients and therefore the quantities computed from them, such as the average and maximum gate performance, depend only on cooperativity, which is independent of the cavity length $L_\mathrm{cav}$ \cite{reiserer2015cavity}. Considering finite pulses would therefore also entail optimizing other system parameters like cavity length or beam splitter delay times $\tau_\mathrm{delay}$.
Similarly, it is worthwhile to investigate speeding up the atom targeting mechanisms such as steering the optical tweezers that move atoms to and from the fiber through optimal control or shortcut to adiabaticity methods \cite{pagano2024optimal,hwang2025fast}.
Employing multiplexing for the photon pulses for parallel gate operation would also provide a significant speedup. Since cavities act as band-pass filters, frequency multiplexing seems to be unsuitable for this purpose, and other methods such as time- or polarization-division multiplexing might need to be used. 
\begin{acknowledgments}
The authors acknowledge funding from JST Moonshot R{\&}D Grant No.~JPMJMS2268.
Circuit diagrams were created using the Quantikz package \cite{kay2023tutorial}.
\end{acknowledgments}

\appendix
\section{Reflection coefficients}
\label{app:reflection_coefficients}
We consider a system of $N$ atoms interacting with the single mode of a cavity at frequency $\omega_c$. Each atom couples to the cavity with a potentially different rate $g_n$ and a level splitting of $\omega_{a,n}$ between the common atomic ground state and excited state $\ket{e_n}$. Spontaneous emission at a rate of $2\gamma$ is included via a non-Hermitian term. 
The cavity couples to the environment via several baths with frequency-independent rates $\sqrt{\kappa_j/\pi}$ and operators fulfilling $[\hat{b}_j(\omega),\hat{b}_j^\dagger(\omega')] = \delta(\omega - \omega')$ \cite{gardiner1985input,cohen2018deterministic}.
The system can be described by the Hamiltonian
\begin{equation}
    \hat{H}_\mathrm{tot} = \hat{H}_\mathrm{sys} + \hat{H}_\mathrm{B} + \hat{H}_\mathrm{int} \; ,
\end{equation}
where the individual components are given by
\begin{subequations}
\begin{align} 
    \hat{H}_\mathrm{sys}/\hbar &=  \omega_c\hat{a}^\dagger\hat{a} + \sum_{n=1}^N\left(\omega_{a,n} - i\gamma\right)\ket{e_n}\bra{e_n}\nonumber\\ & \hphantom{=} + \sum_{n=1}^N g_n\left(\hat{a}^\dagger\ket{g}\bra{e_n} + \ket{e_n}\bra{g}\hat{a}\right) \; ,\\ 
    \hat{H}_\mathrm{B}/\hbar &= \sum_j\int\hspace{-0.25em}d\omega\; \omega\hat{b}_j^\dagger(\omega)\hat{b}_j(\omega) \; ,\\
    \hat{H}_\mathrm{int}/\hbar &= i\sum_{j}\sqrt{\frac{\kappa_j}{\pi}}\int \hspace{-0.25em}d\omega\left(\hat{a}^\dagger\hat{b}_j(\omega) - \hat{b}_j^\dagger(\omega)\hat{a}\right) \; .
\end{align}
Here $\ket{g}:=\ket{q_N\ldots q_2q_1}$ denotes the common atomic ground state in a given qubit configuration.
\end{subequations}
We are interested in the evolution of a single photon impinging onto the cavity and consider a general state containing a single excitation
\begin{equation}
\begin{split}
\ket{\psi(t)} &= \sum_{n=1}^N C_{e,n}(t)\ket{e_n}_a\ket{0}_c\ket{0} + C_{g}(t)\ket{g}_a\ket{1}_c\ket{0}\\ &+ \sum_{j}\ket{g}_a\ket{0}_c\int d\omega f_j(\omega,t)\hat{b}_j^\dagger(\omega)\ket{0}_j \; .
\end{split}
\end{equation}
The coefficients evolve according to the coupled equations
\begin{subequations}
    \begin{align}
        \dot{f}_j(\omega,t) &= -i\omega f_j(\omega,t) + \sqrt{\frac{\kappa_j}{\pi}}C_{g}(t) \; ,\\
        \dot{C}_{g}(t) &= -i\omega_c C_{g}(t) - i\sum_ng_nC_{e,n}(t)\nonumber\\
        &\hphantom{=} - \sum_j\sqrt{\frac{\kappa_j}{\pi}}\int d\omega\;f_j(\omega,t) \; ,\\
        \dot{C}_{e,n}(t) &= -i(\omega_{a,n} -i\gamma)C_{e,n}(t) - ig_nC_{g}(t) \; .
    \end{align}
\end{subequations}
The equation for $f_j(\omega,t)$ is solved by
\begin{equation}
    f_j(\omega,t) = e^{-i\omega t}f_j(\omega,0) + \sqrt{\frac{\kappa_j}{\pi}}\int_0^t ds\;e^{-i\omega(t-s)}C_{g}(s) \; ,
    \label{eq:fj_formal_solution}    
\end{equation}
which, when integrated over the frequency range, yields
\begin{equation}
    \frac{1}{\sqrt{2\pi}}\int d\omega f_j(\omega,t) = \hat{b}_\mathrm{in,j}(t) + \frac{1}{2}\sqrt{2\kappa_j}C_{g}(t) \; ,
    \label{eq:input_relation}
\end{equation}
where we have defined the input pulses for the individual channels
\begin{equation}
    \hat{b}_\mathrm{in,j}(t) = \frac{1}{\sqrt{2\pi}}\int d\omega e^{-i\omega t} f_j(\omega,0) \; .
\end{equation}
Fourier transforming the remaining equations for $C_g$ and $C_e$ and using Eq.~\eqref{eq:input_relation} we obtain their solutions as
\begin{equation}
    C_{e,n}(\omega) = \frac{g_nC_{g}(\omega)}{\omega - \omega_{a,n} + i\gamma} \; ,
\end{equation}
\begin{equation}
    C_g(\omega) = \frac{\sum_j \sqrt{2\kappa_j}\hat{b}_\mathrm{in,j}(\omega)}{i\left(\omega - \omega_c\right) - \sum_j\kappa_j  + \sum_n\frac{g_n^2}{i(\omega - \omega_{a,n}) - \gamma}} \; .
\end{equation}
We are interested in the state of the system at a time $t\rightarrow\infty$ after the photon interacted with the cavity and since further $C_g(t<0)\equiv 0$ we can modify Eq.~\eqref{eq:fj_formal_solution} to obtain
\begin{equation}
\thinmuskip=0mu
\medmuskip=0mu
\begin{split}
    f_j(\omega,t\rightarrow\infty) &= e^{-i\omega t}\left[f_j(\omega,0) + \frac{\sqrt{2\kappa_j}}{2\pi}\int_{-\infty}^{\infty}\hspace{-0.75em}ds\;e^{i\omega s}C_{g}(s)\right]\\
    &= e^{-i\omega t}\left[\hat{b}_\mathrm{in,j}(\omega) + \sqrt{2\kappa_j}C_g(\omega)\right] \; .
\end{split}
\end{equation}
In this limit we also have $C_{g}(t\rightarrow\infty)=0$ and $C_{e,n}(t\rightarrow\infty)=0$ and the final system state then reads $\ket{\psi(t\rightarrow\infty)} = \ket{g}_a\ket{0}_c\ket{p}$ with the photonic component $\ket{p}= \sum_{j}\int d\omega f_j(\omega,t\rightarrow\infty)\hat{b}_j^\dagger(\omega)\ket{0}_j$.

We assume that input into the system comes only from a single channel indexed by $j=0$ and further consider a long-pulse limit in which the duration of the incoming photon pulse is long compared to the timescales determined by $\kappa_j$, $g_n$ and $\gamma$, consisting only of a narrow frequency band centered around $\omega_p$. In this case we have $\hat{b}_\mathrm{in,j} = \delta_{j0}\delta(\omega - \omega_p)$, yielding $\ket{p} = e^{-i\omega_p t}\sum_j f_j(\omega_p)\ket{1}_j$ with
\begin{equation}
    f_j(\omega_p) = \delta_{j0} + \frac{2\sqrt{\kappa_0\kappa_j}}{i\left(\omega_p - \omega_c\right) - \sum_j\kappa_j + \sum_n\frac{g_n^2}{i(\omega_p - \omega_{a,n}) - \gamma}} \; .
\end{equation}
The interaction of the photon with the atoms inside the cavity creates an additional loss channel in which the photon can leave the cavity by scattering off the atoms. This becomes apparent when calculating 
\begin{equation}
    \braket{p|p} = 1 + \frac{4\kappa_0\operatorname{Re}\lbrace g(N,\omega_p)\rbrace}{\tilde{\kappa}^2 + \tilde{\omega}^2} < 1 \; ,
\end{equation}
where we have defined $\tilde{\kappa} = -\kappa + \operatorname{Re}\lbrace g(N,\omega_p)\rbrace$ and $\tilde{\omega} = \omega_p - \omega_c + \operatorname{Im}\lbrace g(N,\omega_p)\rbrace$ with the total loss channel decay rate $\kappa = \sum_j\kappa_j$ and the overall coupling strength function
\begin{equation}
g(N,\omega) = \sum_{n=1}^N\frac{g_n^2}{i\left(\omega - \omega_{a,n}\right) - \gamma} \; . 
\end{equation}
In order to ensure a normalized wave function in agreement with our initial assumption of a single excitation, this additional decay channel has to be accounted for in the photon wave function according to 
\begin{equation}
    \ket{p} \rightarrow \ket{p} + e^{-i\omega_p t}\frac{2\sqrt{\kappa_0\kappa_a}}{i\left(\omega_p - \omega_c\right) - \kappa + g(N,\omega_p)}\ket{p_a}
\end{equation}
with a decay rate 
\begin{equation}
    \kappa_a = - \operatorname{Re}\lbrace g(N,\omega_p)\rbrace = \gamma\sum_{n=1}^N\frac{g_n^2}{\left(\omega_p - \omega_{a,n}\right)^2 + \gamma^2} \; .
\end{equation}

In the physical model studied in the main text we consider three direct cavity decay channels given by reflection, transmission, and scattering at the cavity mirrors with rates $\kappa_r$, $\kappa_t$ and $\kappa_m$ and resulting states denoted as $\ket{p_r}$, $\ket{p_t}$ and $\ket{p_m}$, respectively. Including the additional decay channel from scattering off the atoms with rate $\kappa_a$ and the resulting state $\ket{p_a}$ while neglecting the global phase, the system state finally reads 
\begin{equation}
    \ket{p} = r(\omega_p)\ket{p_r} + t(\omega_p)\ket{p_t} + m(\omega_p)\ket{p_m} + a(\omega_p)\ket{p_a} \; ,
\end{equation}
with the individual amplitudes given by
\begin{subequations}
    \begin{align}
        r(\omega) &= 1 + \frac{2\kappa_r}{i\left(\omega - \omega_c\right) - \kappa + g(N,\omega)} \; ,\\[0.5em]
        t(\omega) &= \frac{2\sqrt{\kappa_r\kappa_t}}{i\left(\omega - \omega_c\right) - \kappa + g(N,\omega)} \; ,\\[0.5em]
        m(\omega) &= \frac{2\sqrt{\kappa_r\kappa_m}}{i\left(\omega - \omega_c\right) - \kappa + g(N,\omega)} \; ,\\[0.5em]
        a(\omega) &= \frac{2\sqrt{\kappa_r\kappa_a}}{i\left(\omega - \omega_c\right) - \kappa + g(N,\omega)} \; .
    \end{align}
\end{subequations}
\section{Calculating the fundamental fiber mode}
\label{app:fiber_mode}
For calculating the fiber coupling strength as a function of atomic distance, we need the cavity field $\mathbf{E}(\mathbf{r})$, which in cylindrical coordinates is given by
\begin{equation}
    \mathbf{E}(\mathbf{r}) = \left(E_r(\mathbf{r}),E_\varphi(\mathbf{r}),E_z(\mathbf{r})\right)^T \; ,
\end{equation}
where the individual field components are obtained by solving Maxwell's equations for a step-index optical fiber with a cladding of refractive index $n_1$ and a core of radius $a$ with refractive index $n_0$. We first define the effective dimensionless wavenumbers $u=a\sqrt{n_1^2k^2 - \beta^2}$ and $w=a\sqrt{\beta^2 - n_0^2k^2}$ for light with wavenumber $k$ propagating through the fiber. 
The propagation constant $\beta$ of the hybrid mode is then calculated from \cite{kien2017higher,okamoto2021fundamentals}
\begin{equation}
    \frac{\beta^2}{k^2} = \left(\frac{n_1^2}{u^2} + \frac{n_0^2}{w^2}\right)\left(\frac{1}{u^2} + \frac{1}{w^2}\right)^{-1}
\end{equation}
for wavenumbers $u$ and $w$ and a fixed integer $l$ that fulfill the relation 
\begin{equation}
\begin{split}
    &l^2\left(\frac{n_1^2}{u^2} + \frac{n_0^2}{w^2}\right)\left(\frac{1}{u^2} + \frac{1}{w^2}\right) =\\[0.5em] &\left(n_1^2\frac{J_l'(u)}{uJ_l(u)} + n_0^2\frac{K_l'(w)}{wK_l(w)}\right)\left(\frac{J_l'(u)}{uJ_l(u)} + \frac{K_l'(w)}{wK_l(w)}\right)
\end{split}
\end{equation}
under the constraint that
\begin{equation}
    u^2 + w^2 = a^2k^2\left(n_1^2 - n_0^2\right)\equiv v^2 \; .
\end{equation}
The quantity $v$ is also known as the fiber volume and $J_l(x)$ and $K_l(x)$ are the Bessel functions of the first kind and modified Bessel functions of the second kind respectively. 
In the following we will only consider the fundamental fiber mode HE$_{11}$ with $l=1$ of an optical nanofiber for which the cladding is the surrounding air with a refractive index of $n_0=1$ and a core with $n_1\approx 1.45$. Throughout the paper we assume a fiber radius of $a=200$ nm and a wavenumber of $k=7.372$ $\mu\mathrm{m}^{-1}$ corresponding to the D2 line of Cesium at a wavelength of $\lambda = 852.3$ nm. 

The unnormalized electric field outside the fiber for $r>1$ (in units of $a$) is given by \cite{yariv1991optical}
\begin{equation}
\thinmuskip=0mu
\medmuskip=0mu
    \begin{pmatrix}
        E_r\\
        E_\varphi\\
        E_z
    \end{pmatrix}=\frac{J_l(u)}{K_l(w)}
    \begin{pmatrix}
        i\frac{a\beta}{w}\left[K_l'(wr) - \frac{ls}{wr}K_l(wr)\right]\\[0.5em]
        -\frac{a\beta}{w}\left[\frac{l}{wr}K_l(wr) + sK_l'(wr)\right]\\[0.5em]
         K_l(wr) 
    \end{pmatrix}e^{i\left(l\varphi -\beta z\right)}
\label{eq:fieber_field_outside}
\end{equation}
and inside the fiber for $r<1$ (in units of $a$) we have
\begin{equation}
    \begin{pmatrix}
        E_r\\
        E_\varphi\\
        E_z
    \end{pmatrix} = 
    \begin{pmatrix}
        -i\frac{a\beta}{u}\left[J_l'(ur) - \frac{ls}{ur}J_l(ur)\right]\\[0.5em]
        \frac{a\beta}{u}\left[\frac{l}{ur}J_l(ur) - sJ_l'(ur)\right]\\[0.5em]
         J_l(ur) 
    \end{pmatrix}e^{i\left(l\varphi -\beta z\right)} \; ,
\label{eq:fieber_field_inside}
\end{equation}
where we have introduced the dimensionless quantity $s$ as an abbreviation for
\begin{equation}
    s = l\left(\frac{1}{u^2} + \frac{1}{w^2}\right)\left(\frac{J_l'(u)}{uJ_l(u)} + \frac{K_l'(w)}{wK_l(w)}\right)^{-1} \; .
\end{equation}
Finally, for plotting the light field according to the coordinate system shown in Fig.~\ref{fig:setup}, we transform back to Cartesian coordinates via the orthogonal matrix
\begin{equation}
    \begin{pmatrix}
        E_x\\
        E_y\\
        E_z
    \end{pmatrix} = 
    \begin{pmatrix}
        \cos\varphi & -\sin\varphi & 0\\
        \sin\varphi & \cos\varphi & 0\\
        0 & 0 & 1
    \end{pmatrix}
    \begin{pmatrix}
        E_r\\
        E_\varphi\\
        E_z
    \end{pmatrix} \; .
\end{equation}
The field in Eqs.~\eqref{eq:fieber_field_outside} and \eqref{eq:fieber_field_inside} describes circularly polarized light. For obtaining quasi-linearly polarized light we need to consider a superposition of left- and right-handed circularly polarized fields according to
\begin{equation}
    \mathbf{E}^\mathrm{lin} = \frac{1}{\sqrt{2}}\left(\mathbf{E}^\mathrm{left}e^{-i\varphi_0} + \mathbf{E}^\mathrm{right}e^{i\varphi_0}\right) \; ,
\end{equation}
where the angle $\varphi_0$ denotes the polarization direction. The left- or right-handedness of the circularly polarized light is determined by the sign of the $E_\varphi$ component and the sign of $l$ in the exponent of $e^{i\left(l\varphi -\beta z\right)}$.
\section{Determining quantum gate error rates}
\label{app:error_rates}
Completely positive maps $\Lambda$ acting on a density matrix $\rho$ such as the quantum channels studied in this work can generally be represented as
\begin{equation}
    \Lambda(\rho) = \sum_m E_m\rho E_m^\dagger
\end{equation}
with some Kraus operators fulfilling $\sum_m E_m^\dagger E_m=\mathbb{I}$ if the map is trace preserving as well \cite{nielsen2010quantum}. In order to calculate the noise bias of a faulty quantum process however, we need to know the error rates $p_j$ in its Pauli channel representation, i.e.~for a single qubit
\begin{equation}
    \Lambda(\rho) = p_0\rho + p_x\hat{X}\rho\hat{X} + p_y\hat{Y}\rho\hat{Y} + p_z\hat{Z}\rho\hat{Z} \, ,
\end{equation}
with $p_0 = 1 - p_x - p_y - p_z$ for a trace preserving process.
For this purpose we first define a twirled version of $\Lambda$ as
\begin{equation}
    \tilde{\Lambda}(\rho) = \frac{1}{K} \sum_{A\in\mathcal{A}} A^{-1}\Lambda\left(A\rho A^{-1}\right)A 
\end{equation}
with a set of $K$ operations $\mathcal{A}=\lbrace A_k\rbrace_{k=1}^K$. If we choose $\mathcal{A}$ as the $N$-qubit Pauli basis, also known as Pauli twirling approximation (PTA), consisting of the $4^N$ distinct operators $\mathcal{A}_N = \lbrace I,X,Y,Z\rbrace^{\otimes N}$, 
the resulting map is diagonal in the Pauli basis \cite{geller2013efficient} 
\begin{equation}
    \tilde{\Lambda}(\rho) = \sum_{A\in A_N} p_a A \rho A^\dagger \, .
\end{equation}
The probabilities $p_A$ are determined by the Kraus operators and fulfill $\sum_{A\in A_N}p_A = 1$ if the map is trace preserving.

From this diagonal form we can determine the error rates of a faulty process for example by adjusting for the target unitary $\hat{U}_\mathrm{ideal}$ and considering $\Lambda(\rho) = \hat{U}_\mathrm{ideal}^\dagger\Lambda(\rho)\hat{U}_\mathrm{ideal}$ instead. We then calculate the Pauli transfer matrix (PTM) defined as
\begin{equation}
    (R_\Lambda)_{ij} = \frac{1}{2^N}\operatorname{Tr}\lbrace \sigma_i\Lambda(\sigma_j)\rbrace 
    \label{eq:ptm_single}
\end{equation}
for a $N$-qubit system and $\sigma_i\in\mathcal{A}_N$. For a Pauli-twirled channel like $\tilde{\Lambda}$ the PTM is diagonal as well and for a single qubit channel we can determine its entries via
\begin{equation}
    (R_{\tilde{\Lambda}})_{\alpha\alpha} = \sum_{\beta,\gamma}p_\beta\left(\theta_{\alpha\beta\gamma}+i\epsilon_{\alpha\beta\gamma}\right)^2
\end{equation} 
by making use of $\sigma_\alpha\sigma_\beta = \left(\theta_{\alpha\beta\gamma}+i\epsilon_{\alpha\beta\gamma}\right)\sigma_\gamma$ and $\operatorname{Tr}(\sigma_\alpha\sigma_\beta) = 2\delta_{\alpha\beta}$, using generalized versions of the Kronecker delta 
\begin{equation}
    \theta_{\alpha\beta\gamma} \equiv 
    \begin{cases}
        1 &\text{one index is 0, the other two equal}\\
        0 &\text{otherwise}
    \end{cases}
\end{equation}
and Levi-Civita tensor
\begin{equation}
    \epsilon_{\alpha\beta\gamma} \equiv 
    \begin{cases}
        1 &\alpha\beta\gamma\in\lbrace 123,231,312\rbrace\\
        -1 &\alpha\beta\gamma\in\lbrace 321,213,132\rbrace\\
        0 &\text{repeated indices, or any index is 0}
    \end{cases}
\end{equation}
that include the identity matrix $\sigma_0$ \cite{gamel2016entangled}. We can also express Eq.~\eqref{eq:ptm_single} in the following form with some matrix $A_1$
\begin{equation}
    \begin{pmatrix}
        R_{11}\\ R_{22}\\ R_{33}\\ R_{44}\\
    \end{pmatrix}
    \equiv A_1
        \begin{pmatrix}
        p_0\\ p_x\\ p_y\\ p_z\\
    \end{pmatrix}=
    \begin{pmatrix}
        1 & 1 & 1 & 1\\
        1 & 1 & -1 & -1\\
        1 & -1 & 1 & -1\\
        1 & -1 & -1 & 1\\
    \end{pmatrix}
    \begin{pmatrix}
        p_0\\ p_x\\ p_y\\ p_z\\
    \end{pmatrix} \; ,
\end{equation}
which is inverted by $A_1^{-1} = A_1/4$ to determine the Pauli channel error rates from the PTM diagonal according to
\begin{equation}
    \begin{pmatrix}
        p_0\\ p_x\\ p_y\\ p_z\\
    \end{pmatrix}=\frac{1}{4}
    \begin{pmatrix}
        1 & 1 & 1 & 1\\
        1 & 1 & -1 & -1\\
        1 & -1 & 1 & -1\\
        1 & -1 & -1 & 1\\
    \end{pmatrix}
    \begin{pmatrix}
        R_{11}\\ R_{22}\\ R_{33}\\ R_{44}\\
    \end{pmatrix} \, .
\end{equation}

For a two-qubit channel, using similar relations $\operatorname{Tr}\left(D_{\alpha\beta}D_{\gamma\delta}\right)=4\delta_{\alpha\gamma}\delta_{\beta\delta}$ and
\begin{equation}
    D_{\alpha\beta}D_{\gamma\delta} = \left(\theta_{\alpha\gamma\mu}+i\epsilon_{\alpha\gamma\mu}\right)\left(\theta_{\beta\delta\nu}+i\epsilon_{\beta\delta\nu}\right)D_{\mu\nu}
\end{equation}
for the Dirac matrices $D_{\mu\nu} = \sigma_\mu\otimes\sigma_\nu$, we can analogously determine the PTM diagonal entries according to 
\begin{equation}
    \thinmuskip=0mu
    \medmuskip=0mu
    (R_\Lambda)_{(\alpha\beta)(\alpha\beta)} = 
\sum_{\mu,\nu}p_{\mu\nu}\sum_{\gamma,\delta}\left(\theta_{\alpha\mu\gamma}+i\epsilon_{\alpha\mu\gamma}\right)^2\left(\theta_{\beta\nu\delta}+i\epsilon_{\beta\nu\delta}\right)^2,
\end{equation} 
where $(\alpha\beta)$ is a bipartite index with $\alpha=1,\ldots,4$ and $\beta=1,\ldots,4$.

The resulting matrix and its inverse are simply given by $A_2 = A_1\otimes A_1$ and $A_2^{-1} = A_1^{-1}\otimes A_1^{-1}$ respectively.
From the error rates determined in this way, we can then further calculate the noise bias of a certain operation according to Eq.~\eqref{eq:bias_definition} in the main text.

Similarly, when studying systems with $N>2$ we use $A_N=A_1^{\otimes N}$ and trace out all qubits except the control and target of the CZ gate. For instance, in the $N=4$ case shown in Fig.~\ref{fig:pauli_error_rates_N4_without_addressing} in the main text, we have
\begin{equation}
    p_{ij} = \sum_{kl} p_{ijkl}
\end{equation}
for the relevant probabilities $p_{ij}$ and $k,l\in\lbrace I,X,Y,Z\rbrace$.
\end{document}